\documentclass[twocolumn,superscriptaddress,showpacs,aps,prb,amsmath,amssymb]{revtex4}
\usepackage{bm}
\usepackage{graphicx}
\usepackage{color}
\newcommand{\B}{\text{\scriptsize bath}}
\newcommand{\s}{\text{\scriptsize sys}}

\newcommand{\T}{{\rm total}}
\newcommand{\w}{\omega}
\newcommand{\dg}{\dagger}
\newcommand{\ti}{\tilde}

\newcommand{\nl}{\nonumber \\}
\newcommand{\la}{\langle}
\newcommand{\ra}{\rangle}

\newcommand{\Sec}[1]{Sec.\,\ref{#1}}

\newcommand{\be}{\begin{equation}}
\newcommand{\ee}{\end{equation}}
\newcommand{\bsube}{\begin{subequations}}
\newcommand{\esube}{\end{subequations}}
\newcommand{\Eq}[1]{Eq.\,(\ref{#1})}
\newcommand{\Eqs}[1]{Eqs.\,(\ref{#1})}
\newcommand{\Fig}[1]{Fig.\,\ref{#1}}

\newcommand{\bfL}  {\mbox{\boldmath${\cal L}$}}
\newcommand{\bfV}  {\mbox{\boldmath${\cal V}$}}
\newcommand{\bfG}  {\mbox{\boldmath${\cal G}$}}
\newcommand{\bfone}  {\mbox{\boldmath${\cal I}$}}
\newcommand{\bfS}  {\mbox{\boldmath${\cal S}$}}
\newcommand{\comments}[1]{}

\begin{document}

\title{Hierarchical Liouville-space approach to nonequilibrium
dynamical properties of quantum impurity systems}

\author{Shikuan Wang}
\affiliation{Department of Chemistry, Hong Kong University of
 Science and Technology, Kowloon, Hong Kong, China}

\author{Xiao Zheng} \email{xz58@ustc.edu.cn}
\affiliation{Hefei National Laboratory for Physical Sciences at the
 Microscale, University of Science and Technology of China, Hefei,
 Anhui 230026, China}

\author{Jinshuang Jin}
\affiliation{Department of Physics, Hangzhou Normal University,
Hangzhou, Zhejiang 310036, China
}

\author{YiJing Yan}\email{yyan@ust.hk}
\affiliation{Department of Chemistry, Hong Kong University of
 Science and Technology, Kowloon, Hong Kong, China}
\affiliation{Hefei National Laboratory for Physical Sciences at the
 Microscale, University of Science and Technology of China, Hefei,
 Anhui 230026, China}

%
\date{Submitted on October~24, 2012; Resubmitted on July~10, 2013}

\begin{abstract}
We propose a hierarchical dynamics approach for evaluation of nonequilibrium
dynamic response properties of quantum impurity systems. It is based on a
hierarchical equations of motion formalism, in conjunction with a linear response
theory established in the hierarchical Liouville space.
This provides an accurate and universal tool for characterization of a
variety of response and correlation functions of local impurities, as
well as transport related response properties.
The practicality of our proposed approach is demonstrated via the
evaluation of various dynamical properties of a single-impurity
Anderson model. These include the impurity spectral density, local
charge fluctuation, local magnetic susceptibility, and current-voltage
admittance, in both equilibrium and nonequilibrium situations. The
numerical results are considered to be quantitatively accurate, as long
as they converge quantitatively with respect to the truncation of the
hierarchy.
\end{abstract}

\maketitle

\section{Introduction}

\label{thintro}

Recent advances in fabrication, manipulation and measurement of artificial
quantum impurity systems such as quantum dots have led to a resurgence of
interest of nanostructures in both experiment and theory. A favorable feature
of these nanostructures is the outstanding tunability of device parameters.
Understanding the dynamical properties of quantum impurity systems is of
fundamental importance for the development of solid-state quantum information
processing\cite{Elz04431, Kop06766, Han081043} and single-electron devices.
\cite{Gab06499, Fev071169}
Moveover, quantum impurity models serve as essential theoretical tools, covering
a broad range of important physical systems. For instance, the Hubbard lattice
model can be mapped onto the Anderson impurity model via a self-consistent
dynamical mean field theory.\cite{Met89324, Geo926479, Geo9613} Besides the
strong electron-electron (\emph{e-e}) interactions, local impurities are also
subject to interactions with itinerant electrons in surrounding bulk materials,
which serve as the electron reservoir as well as thermal bath.
The interplay between the local \emph{e-e} interactions and nonlocal transfer
coupling gives rise to a variety of intriguing phenomena of prominent many-body
nature, such as Kondo effect,\cite{Cro98540, Gol98156, Bul08395} Mott metal-insulator
transition,\cite{Geo921240, Ima981039, Bul01045103} and high-temperature
superconductivity.\cite{Eme872794, Yan031, Mai05237001}

Characterizing the system responses to external perturbation of experimental
relevance is of fundamental significance in understanding the intrinsic
properties of quantum impurity systems and their potential applications.
For instance, the magnetic susceptibility of an impurity system reflects
the redistribution of electron spin under an applied magnetic field, and
its investigation may have important implications for fields such as spintronics.

For the accurate characterization of dynamical properties of the
impurity such as the impurity spectral function and dynamical
charge/magnetic susceptibility, a variety of nonperturbative numerical
approaches have been developed, such as numerical renormalization group
method,\cite{Wil75773, Hof001508, Bul08395} density matrix
renormalization group approach,\cite{Whi922863, Jec02045114, Nis04613}
and quantum Monte Carlo method. \cite{Hir862521,Sil902380, Gub916011,
Gul11349}
While most of work has focused on equilibrium properties, the accurate
characterization of nonequilibrium dynamical properties has remained very
challenging.

In many experimental setups,\cite{Gol98156, Cro98540} artificial quantum
impurity systems attached to electron reservoirs are subject to applied bias
voltages. This stimulates the experimental and theoretical exploration of
nonequilibrium processes in quantum impurity systems. A variety of interesting
physical phenomena have been observed, which originate from the interplay between
strong electron correlation and nonequilibrium dissipation.\cite{Doy06245326,
Meh08086804, Bou08140601, Chu09216803}

In the past few years, a number of nonperturbative theoretical approaches
have been devised to treat systems away from equilibrium. These include
the time-dependent numerical renormalization group method,\cite{Cos973003, And05196801,
And08066804} time-dependent density matrix renormalization group method,\cite{Whi04076401}
nonequilibrium functional renormalization group,\cite{Jak07150603, Gez07045324}
quantum Monte Carlo method,\cite{Han99236808, Wer09035320, Sch09153302} iterative
real-time path integral approach,\cite{Wei08195316, Seg10205323} and nonequilibrium
Bethe ansatz.\cite{Kon01236801, Meh06216802, Cha11195314}
Despite the progress made, quantitative accuracy is not guaranteed for the
resulted nonequilibrium properties, because of the various simplifications and
approximations involved in these approaches. Therefore, an accurate and universal
approach which is capable of addressing nonequilibrium situations is highly
desirable.

In this work we propose a hierarchical dynamics approach for the characterization
of nonequilibrium response of local impurities to external fields.
A general hierarchical equations of motion (HEOM) approach has been developed,
\cite{Jin08234703, Zhe09164708, Zhe121129} which describes the reduced dynamics
of open dissipative systems under arbitrary time-dependent external fields. The
HEOM theory resolves the combined effects of \emph{e-e} interactions, impurity-reservoir
dissipation, and non-Markovian memory in a nonperturbative manner.
In the framework of HEOM, the nonequilibrium dynamics are treated by following the
same numerical procedures as in equilibrium situations. The HEOM theory is in principle
exact for an arbitrary equilibrium or nonequilibrium system, provided that the full hierarchy
inclusive of infinite levels are taken into account.\cite{Jin08234703} In practice, the
hierarchy needs to be truncated at a finite level for numerical tractability. The convergence
of calculation results with respect to the truncation level should be carefully examined.
Once the convergence is achieved, the numerical outcome is considered to reach quantitative
accuracy for systems in both equilibrium and nonequilibrium situations.

It has been demonstrated that the HEOM approach leads to an accurate
and universal characterization of strong electron correlation effects
in quantum impurity systems, and treats the equilibrium and
nonequilibrium scenarios in a unified manner. For the equilibrium
properties of Anderson model systems, the HEOM approach achieves the
same level of accuracy as the latest state-of-the-art NRG
method.\cite{Li12266403}
In particular, the universal Kondo scaling of zero-bias conductance and the logarithmic
Kondo spectral tail have been reproduced quantitatively.
For systems out of equilibrium, numerical calculations achieving
quantitative accuracy remain very scarce. One of the rare cases where
numerically exact solution is available is the dynamic current response
of a noninteracting quantum dot to a step-pulse voltage.
\cite{Ste04195318,Mac06085324} This has been precisely reproduced by
the HEOM approach. \cite{Zhe08184112} However, there are very few
calculations at the level of quantitative accuracy for systems
involving strong \emph{e-e} interactions, since most of the existing
methods involve intrinsic approximations. Based on the HEOM formalism,
quantitative accuracy should be achieved once the numerical convergence
with respect to the truncation level of hierarchy is reached.

There are two schemes to evaluate the response properties of quantum
impurity systems in the framework of HEOM: (\emph{i}) Calculate
relevant system correlation/response functions based on a linear
response theory constructed in the HEOM Liouville space;\cite{Li12266403}
and (\emph{ii}) solve the EOM for a hierarchical set of density
operators to obtain the transient reduced dynamics of system in
response to time-dependent external perturbation, followed by a finite
difference analysis. These two schemes are completely equivalent in the
linear response regime, as have been verified numerically.
In previous studies, we had employed the above second scheme to
evaluate the dynamic admittance (frequency-dependent electric current
in response to external voltage applied to coupling electron
reservoirs) of quantum dot systems, which had led to the identification
of several interesting phenomena, including dynamic Coulomb
blockade\cite{Zhe08093016} and dynamic Kondo transition,\cite{Zhe09164708}
and photon-phonon-assisted transport.\cite{Jia12245427}

In this work, we will elaborate the above first scheme of HEOM approach.
The external perturbation may associate with an arbitrary operator in
the impurities subspace, or originates from a homogeneous shift of
electrostatic potential (and hence the chemical potential) of electron
reservoir. The detailed numerical procedures will be exemplified through the
evaluation of a variety of response properties of a single-impurity
Anderson model, including the impurity spectral density function, local
charge fluctuation spectrum, local magnetic susceptibility, and dynamic
admittance.

The remainder of paper is organized as follows. We will first give a
brief introduction on the HEOM method in \Sec{thheom}.
In \Sec{ththeo} we will elaborate the establishment of a linear
response theory in the HEOM Liouville space. Calculation on system
correlation/response functions which are directly relevant to the
response properties of primary interest will be discussed in detail.
We will then provide numerical demonstrations for the evaluation of
various dynamical properties in \Sec{thnum}. Finally, the concluding
remarks will be given in \Sec{summary}.

\section{A real-time dynamics theory for nonequilibrium impurity systems}
\label{thheom}

\subsection{Prelude}
\label{thheomA}

  Consider a quantum impurity system in contact with two electron reservoirs,
denoted as the $\alpha=$ L and R reservoirs, under the bias
voltage $V=\mu_{\rm L}-\mu_{\rm R}$.
 The total Hamiltonian
of the composite system assumes the form of
\begin{align}
\label{Htotal}
H_{\T}&=H_{\s}+\sum_{\alpha k} (\epsilon_{\alpha k} + \mu_\alpha)\,
\hat d^{\dg}_{\alpha k}\hat d_{\alpha k}
\nl& \quad
+\sum_{\alpha\mu k}\left(t_{\alpha k\mu}\hat d^{\dg}_{\alpha k}\hat a_{\mu}
+ {\rm H.c.} \right).
\end{align}
The impurity system Hamiltonian $H_{\s}$ is rather general, including
many-particle interactions and external field coupling. Its second
quantization form is given in terms of electron creation and
annihilation operators, $\hat a^\dg_{\mu}\equiv \hat a^+_{\mu}$ and
$\hat a_{\mu}\equiv \hat a^-_{\mu}$, which are associated with the
system spin-state $\mu$.
The reservoirs are modeled by a noninteracting Hamiltonian; see the
second term on the right-hand side (rhs) of \Eq{Htotal}, where $\hat
d^{\dg}_{\alpha k}$ ($\hat d_{\alpha k}$) and $\epsilon_{\alpha k}$ are
the creation (annihilation) operator and energy of single-electron
state $|k\rangle$ electron of $\alpha$-reservoir, respectively. While
the equilibrium chemical potential of total system is set to be
$\mu^{\rm eq}_{\alpha}=0$, the reservoir states are subject to a
homogeneous shift, $\mu_\alpha$, under applied voltages.
The last term on the rhs of \Eq{Htotal} is in a standard transfer
coupling form, which is responsible for the dissipative interactions
between the system and itinerary electrons of reservoirs. It can be
recast as $H'=\sum_{\alpha\mu}(\hat f^{+}_{\alpha\mu}
 \hat a^{-}_{\mu}+\hat a^{+}_{\mu}\hat f^{-}_{\alpha\mu})$,
where $\hat f^{+}_{\alpha\mu}\equiv \sum_{k}t_{\alpha k\mu}
\hat d^{\dg}_{\alpha k}=\big(\hat f^{-}_{\alpha\mu}\big)^{\dg}$.
Throughout this paper we adopt the atomic unit $e=\hbar=1$ and
denote $\beta=1/(k_B T)$, with $k_B$ being the Boltzmann constant
and $T$ the temperature of electron reservoirs. Introduce also the
sign variables, $\sigma=+/-$ and $\bar\sigma \equiv -\sigma$
the opposite sign of $\sigma$.

The $\alpha$-reservoir is characterized by the spectral density
$J_{\alpha \mu\nu}(\omega) \equiv \pi \sum_{k}t^\ast_{\alpha k\mu}
t_{\alpha k\nu} \delta(\omega - \epsilon_{\alpha k})$.
It influences the dynamics of reduced system through the reservoir
correlation functions $\tilde{C}^{\sigma;{\rm st}}_{\alpha\mu\nu}
(t-\tau)\equiv \la\hat f^{\sigma}_{\alpha\mu}(t)
  \hat f^{\bar\sigma}_{\alpha \nu}(\tau)\ra_{\alpha}$,
Here, $\la(\cdot)\ra_{\alpha}\equiv {\rm
tr}_{\alpha}\big[(\cdot)\,e^{-\beta H_{\alpha}}\big] /{\rm
tr}_{\alpha}(e^{-\beta H_{\alpha}})$ and $\hat
f^{\sigma}_{\alpha\mu}(t)\equiv  e^{iH_{\alpha}t}\hat
f^{\sigma}_{\alpha\mu} e^{-iH_{\alpha}t}$, with $H_\alpha$ being the
Hamiltonian of $\alpha$-reservoir.
The superscript ``st'' highlights the stationary feature of the
nonequilibrium correlation function, under a constant $\mu_{\alpha}$.
It is related to the reservoir spectral density,
$J_{\alpha\mu\nu}(\omega)\equiv J^{-}_{\alpha\mu\nu}(\omega)\equiv
J^{+}_{\alpha\nu\mu}(\omega)$, via the fluctuation-dissipation theorem:
\cite{Jin08234703}
\be\label{FDT}
\tilde{C}^{\sigma;{\rm st}}_{\alpha\mu\nu}(t)
=\int_{-\infty}^{\infty}\! d\omega
\frac{ e^{\sigma i\omega t}J^{\sigma}_{\alpha\mu\nu}(\omega-\mu_{\alpha})
     }{1+e^{\sigma\beta(\omega-\mu_{\alpha})}}.
\ee
Physically, $\tilde{C}^{\sigma;{\rm st}}_{\alpha\mu\nu}(t)$,
with $\sigma=+$ or $-$,
describes the processes of electron
tunneling from the $\alpha$-reservoir into the
specified system coherent state
or the reverse events, respectively.

 We will be interested in nonequilibrium dynamic responses
to a time-dependent external field acting on either the local
system or the reservoirs. For the latter case, we include a
time-dependent shift in chemical potential $\delta\Delta_{\alpha}(t)$,
on top of the constant $\mu_{\alpha}$, to the $\alpha$-reservoir. Its
effect can be described by rigid homogeneous shifts for the reservoir
conduction bands, resulting in the nonstationary reservoir correlation
functions of
\be\label{CT}
C^{\sigma}_{\alpha\mu\nu}(t,\tau)
=\exp\left[\sigma i\!\int_{\tau}^t\! {\rm d}t'\, \delta\Delta_{\alpha}(t')\right]
\tilde{C}^{\sigma;{\rm st}}_{\alpha\mu\nu}(t-\tau).
\ee
This is the generalization of
$\tilde{C}^{\sigma;{\rm st}}_{\alpha\mu\nu}(t)=
 e^{\sigma i\mu_{\alpha} t} \tilde{C}^{\sigma;{\rm eq}}_{\alpha\mu\nu}(t)$,
as inferred from \Eq{FDT},
with the equilibrium counterpart being of $\mu^{\rm eq}_{\alpha}=0$.
In the following, we focus on the situation of diagonal reservoir
correlation, \emph{i.e.}, $J^{\sigma}_{\alpha\mu\nu}(\omega)=
J^{\sigma}_{\alpha\mu\mu}(\omega)\,\delta_{\mu\nu}$, and
$\tilde{C}^{\sigma;{\rm st}}_{\alpha\mu\nu}(t)
 =\tilde{C}^{\sigma;{\rm st}}_{\alpha\mu\mu}(t)\,\delta_{\mu\nu}$.
 In constructing closed HEOM,\cite{Jin08234703} we should
expand $\tilde{C}^{\sigma;{\rm st}}_{\alpha\mu\mu}(t)$
in a finite exponential series,
\be\label{CTexp}
\tilde{C}^{\sigma;{\rm st}}_{\alpha\mu\mu}(t) \simeq \sum_{m=1}^M
\eta^{\sigma}_{\alpha\mu m}e^{-\gamma^{\sigma}_{\alpha\mu m} t}.
\ee
Involved are a total number of $M=N'+N$
poles from the reservoir spectral density
and the Fermi function in the contour integration
evaluation of \Eq{FDT}.
Various sum-over-poles schemes
have been developed, including the Matsubara spectrum decomposition scheme,\cite{Jin08234703}
a hybrid spectrum decomposition and frequency dispersion scheme,\cite{Zhe09164708}
the partial fractional decomposition scheme,\cite{Cro09073102}
and the Pad\'{e} spectrum decomposition (PSD) scheme,\cite{Hu10101106, Hu11244106}
with the primary focus on the Fermi function.
To our knowledge, the PSD scheme has the best
performance until now. We will come back to
this issue later; see the remark-(6)
in \Sec{thheomB}.
In the present work we use
the $[N\!-\!1/N]$ PSD scheme.\cite{Hu10101106, Hu11244106}
It leads to a minimum $M=N'+N$ in the exponential
expansion of \Eq{CTexp} and thus an optimal HEOM construction.\cite{Jin08234703}

 The exponential expansion form
of the reservoir correlation function in \Eq{CTexp}
dictates the explicit expressions for the HEOM formalism.\cite{Jin08234703}
For bookkeeping we introduce the abbreviated index
$j=\{\sigma\alpha\mu m\}$ for $\gamma_j\equiv \gamma^{\sigma}_{\alpha\mu m}$ and
so on, or $j=\{\sigma\mu\}$ for $\hat a_j \equiv \hat a^{\sigma}_{\mu}$.
Denote also $\bar j=\{\bar\sigma\alpha\mu m\}$ or $\{\bar\sigma\mu\}$ whenever
appropriate, with $\bar\sigma$ being the opposite
sign of $\sigma=+$ or $-$.
The dynamical variables in HEOM are a set
of auxiliary density operators (ADOs), $\{\rho^{(n)}_{j_1\cdots j_n}(t);
n=0,1,\cdots,L\}$, with $L$ being the terminal or truncated tier of hierarchy.
The zeroth-tier ADO is set to be the reduced system
density matrix, $\rho^{(0)}(t)\equiv \rho(t) \equiv {\rm tr}_{\B}\,[\rho_{\T}(t)]$,
\emph{i.e.}, the trace of the total system and bath composite density matrix
over reservoir bath degrees of freedom.

\subsection{Hierarchical equations of motion formalism}
\label{thheomB}

The HEOM formalism has been constructed from the Feynman--Vernon
influence functional path integral theory, together with the Grassmann algebra.\cite{Jin08234703}
The initial system-bath decoupling used for expressing explicitly the
influence functional is set at the infinite past.
It does not introduce any approximation for
the characterization of any realistic physical process starting
from a stationary state, which is defined via the HEOM that
includes the coherence between the system and bath.
The detailed construction of HEOM is referred to Ref.~\onlinecite{Jin08234703}.
Here we just briefly introduce the HEOM formalism and
discuss some of its key features.

 The final HEOM formalism reads\cite{Jin08234703}
\begin{align}\label{HEOM}
\dot\rho^{(n)}_{j_1\cdots j_n} =&-\big[i{\cal L}(t)
+\gamma^{(n)}_{j_1\cdots j_n}\!(t)\big]\rho^{(n)}_{j_1\cdots j_n}
-i \sideset{}{'}\sum_{j}{\cal A}_{\bar j}\, \rho^{(n+1)}_{j_1\cdots j_nj}
\nl &
-i \sum_{r=1}^{n}(-)^{n-r}\, {\cal C}_{j_r}\,
\rho^{(n-1)}_{j_1\cdots j_{r-1}j_{r+1}\cdots j_n}\, .
\end{align}
The boundary conditions are $\gamma^{(0)}
=\rho^{(-1)}=0$, together with a truncation by setting all $\rho^{(n>L)}=0$.
The initial conditions to \Eq{HEOM}
will be specified in conjunction with the evaluation of various response
and correlation functions in \Sec{ththeo}.

The time-dependent damping parameter $\gamma^{(n)}_{j_1\cdots j_n}\!(t)$
in \Eq{HEOM}
collects the exponents of nonstationary reservoir correlation function
[\emph{cf.}~\Eqs{CT} and (\ref{CTexp})]:
\be\label{gammaSum}
\gamma^{(n)}_{j_1\cdots j_n}\!(t)=\sum_{r=1}^{n}
\big[\gamma_{j_r}-\sigma i \delta\Delta_{\alpha}(t)\big]_{\sigma,\alpha\in j_r}.
\ee
This expression has been used directly in
the HEOM evaluation of transient current dynamical properties
under the influence of arbitrary time-dependent chemical
potentials applied to electrode leads.\cite{Zhe08184112,Zhe08093016,Zhe09124508,Zhe09164708}
Note that $\gamma_{j}\equiv \gamma^{\sigma}_{\alpha\mu m}=
\gamma^{\sigma;{\rm eq}}_{\alpha\mu m}-\sigma i \mu_{\alpha}$.
In \Sec{ththeo3}, we will treat $\delta\Delta_{\alpha}(t)$ as perturbation
and derive the corresponding linear response theory formulations
for various transport current related properties under nonequilibrium
($\mu_{\alpha}\neq 0$) conditions.

  To evaluate nonequilibrium correlation functions of local system
via linear response theory (\emph{cf.}~\Sec{ththeo2}),
the time-dependent reduced system Liouvillian in \Eq{HEOM}
is assumed formally the form of
\be\label{calLt}
 \mathcal{L}(t)=\mathcal{L}_{s}+\delta\mathcal{L}(t).
\ee
Here, $\mathcal{L}_s\,\cdot\,\equiv [H_{\s},\,\cdot\,]$ remains
the commutator form involving two $H_{\s}$-actions onto
the bra and ket sides individually.
However, the time-dependent perturbation $\delta{\cal L}(t)$
may act only on one side, in line with the
HEOM expressions for local system correlation functions.%
\cite{Mo05084115,Zhu115678,Xu11497,Xu13024106,Li12266403}

 Other features of HEOM and remarks, covering both the theoretical formulation
and numerical implementation aspects, are summarized as follows.

(1) The Fermi-Grassmannian properties:
(\emph{i}) All $j$-indexes in a nonzero $n^{\rm th}$-tier ADO,
$\rho^{(n)}_{j_1\cdots j_n}$, must be distinct.
Swap in any two of them leads to a minus sign,
such as $\rho^{(2)}_{j_2j_1}=-\rho^{(2)}_{j_1j_2}$.
In line with this property, the sum of the
tier-up dependence in \Eq{HEOM} runs only over $j\not\in \{j_1,\cdots,j_n\}$;
(\emph{ii}) Involved in \Eq{HEOM} are also ${\cal A}_{\bar j}\equiv {\cal A}^{\bar\sigma}_{\mu}$
and ${\cal C}_j\equiv {\cal C}^{\sigma}_{\alpha\mu m}$. They are Grassmann superoperators,
defined via their actions on an arbitrary operator of fermionic or bosonic
(bi-fermion) nature,
$\hat O^{\text{\tiny F}}$ or $\hat O^{\text{\tiny B}}$, by
\be\label{calAC}
\begin{split}
{\cal A}_{\bar j} \hat O^{\text{\tiny F/B}}
&\equiv \hat a_{\bar j}\hat O^{\text{\tiny F/B}}
\mp \hat O^{\text{\tiny F/B}}\hat a_{\bar j} \, ,
\\
{\cal C}_{j} \hat O^{\text{\tiny F/B}}
&\equiv \eta_j\hat a_j\hat O^{\text{\tiny F/B}}
\pm \eta_{\bar j}^{\ast}\hat O^{\text{\tiny F/B}}\hat a_j \, .
\end{split}
\ee
In particular, even-tier ADOs are bosonic, while odd-tier ones are fermionic.
The case of opposite parity would also appear in conjunction with
applications; see comments following \Eq{tibmrho_init1}.

(2) Physical contents of ADOs:
While the zero-tier ADO is the reduced density matrix,
\emph{i.e.}, $\rho^{(0)}(t) = \rho(t)$,
the first-tier ADOs, $\rho^{(1)}_j\equiv \rho^{\sigma}_{\alpha\mu m}$,
are related to the electric current through the interface between the system
and $\alpha$-reservoir, $I_\alpha(t)$, as follows,
\be\label{curr_t}
  I_{\alpha}(t)= - 2\,{\rm Im} \sum_{\mu m}
{\rm Tr}\left[\hat{a}^{+}_{\mu}\rho^{-}_{\alpha\mu m}(t)\right].
\ee
Moreover, we have
$\sum_m \rho^{\sigma}_{\alpha\mu m}(t)=
{\rm tr}_{\B}[\hat f^{\sigma}_{\alpha \mu}(t)\rho_{\T}(t)]$,
and can further relate
${\rm tr}_{\B}[\hat f^{\sigma}_{\alpha\mu}(t)
\hat f^{\sigma'}_{\alpha'\mu'}(t)\rho_{\T}(t)]$ to the second-tier ADOs,
and so on.
Note that $\hat f^{+}_{\alpha\mu}(t)\equiv e^{iH_{\alpha}t}
\big(\sum_{k}t_{\alpha k\mu}
\hat d^{\dg}_{\alpha k}\big)e^{-iH_{\alpha}t}
=\big[\hat f^{-}_{\alpha\mu}(t)\big]^{\dg}$
are defined in the bath-space only.
Apparently, the Fermi-Grassmannian properties in remark-(1) above are rooted at the
fermionic nature of individual $\{\hat f^{\sigma}_{\alpha\mu}\}$.

(3)  Hermitian property: The ADOs satisfy the
 Hermitian relation of $\big[\rho^{(n)}_{j_1\cdots j_n}(t)\big]^{\dg}
 =\rho^{(n)}_{{\bar j}_n\cdots {\bar j}_1}(t)$,
 whenever the perturbed $i\delta{\cal L}(t)$ action
 assumes Hermitian; see the comments
 following \Eq{calLt}.

(4) Nonperturbative nature: The HEOM construction treats properly
the combined effects of system-bath coupling strength,
Coulomb interaction, and bath memory time scales,
as inferred from the following observations.
(\emph{i}) For noninteracting electronic systems, the coupling hierarchy stops at
second tier level ($L=2$) without approximation;\cite{Jin08234703}
(\emph{ii}) HEOM is of finite support, containing in general only a finite number of ADOs.
 Let $K$ be the number of all distinct $j$-indexes.
Such a number draws the maximum tier level
$L_{\text{max}}=K$, at which the HEOM formalism ultimately terminates.
The total number of ADOs, up to the truncated tier level $L$,
is $\sum^{L}_{n=0}\frac{K!}{n!(K-n)!}\leq 2^{K}$, as $L\leq L_{\text{max}}=K$;
(\emph{iii}) The hierarchy resolves collectively the memory contents, as decomposed in
 the exponential expansion of bath correlation functions of \Eq{CTexp}.
 It goes with the observation that an individual ADO, $\rho^{(n)}_{j_1\cdots j_n}$,
 is associated with the collective damping constant
 Re\,$\gamma^{(n)}_{j_1\cdots j_n}$ in \Eq{gammaSum}.
 Meanwhile $\rho^{(n)}_{j_1\cdots j_n}$ has the leading $(2n)^{\rm th}$-order
 in the overall system-bath coupling strength.
 One may define proper non-Marvokianicity parameters to
 determine in advance the numerical importance
 of individual ADOs;\cite{Xu05041103,Shi09084105,Xu13024106}
(\emph{iv}) Convergency tests by far -- For quantum
impurity systems with nonzero e-e interactions, calculations often converge
rapidly and uniformly with the increasing truncation level $L$.
Quantitatively accurate results are usually achieved at a relatively low
value of $L$.

(5) Nonequilibrium versus equilibrium:
 The HEOM formalism presented earlier provides a unified approach
to equilibrium, nonequilibrium, time-dependent and
time-independent situations.
In general, the number $K$ of distinct ADO indexes
amounts to $K=2N_{\alpha}N_{\mu}M$, as inferred from
\Eq{CTexp}, with $N_{\mu}$ being the number
spin-orbitals of system in direct contact to leads.
The factor 2 accounts for the two choices
of the sign variable $\sigma$,
while $N_{\alpha}=2$ for the distinct  $\alpha=$ L and R leads.
Interestingly, in the equilibrium
case, together with the $J_{\rm L}(\omega)\propto  J_{\rm R}(\omega)$
condition, one can merger all leads into a single lead to have the
reduced $K=2N_{\mu}M$.
The resulting equilibrium HEOM formalism that contains no longer the
$\alpha$-index can therefore be evaluated at
the considerably reduced computational cost.

(6) Control of accuracy and efficiency: The bath correlation function
  in exponential expansion of \Eq{CTexp} dictates the accuracy
  and efficiency of the HEOM approach.
(\emph{i}) The accuracy in the exponential expansion of \Eq{CTexp}
 is found to be directly transferable to the accuracy
 of HEOM. In other words, HEOM is exact as long as the expansion is exact;
(\emph{ii}) The expansion of \Eq{CTexp} is uniformly convergent,
and becomes exact when $M$ goes to infinity,
for any realistic bath spectral density with
finite bandwidth at finite temperature ($T \neq 0$);
(\emph{iii}) The $[N\!-\!1/N]$ PSD scheme adopted in this work
is considered to be the best among all possible sum-over-poles
expansion of Fermi function.\cite{Hu10101106,Hu11244106,Bak96}
In particular it is dramatically superior over the commonly
used Matsubara expansion expression.
The PSD scheme leads to the optimal HEOM, with a minimum $K$-space size,
for either equilibrium or nonequilibrium case, as discussed in
remark-(5) above.

(7) Computational cost: The CPU time and memory space required for
HEOM calculations are rather insensitive
to the Coulomb coupling strength and to the
equilibrium versus nonequilibrium and
time-dependent and  time-independent types
of evaluations.
However, it grows exponentially as the temperature $T\rightarrow 0$,
with respect to system-bath hybridization strength,
due to the significant increase of both the
converged $K$-space and $L$-space sizes.

 To conclude, HEOM is an accurate and versatile tool,
capable of universal characterizations
of real-time dynamics in quantum impurity systems,
in both equilibrium and nonequilibrium cases.
These remarkable features have been demonstrated
recently in several complex quantum
impurity systems,\cite{Li12266403}
with the focus mainly on equilibrium properties.
The HEOM approach is also very efficient. Calculations often converge
rapidly and uniformly with the increasing truncation level $L$.
Quantitatively accurate results are usually achieved at
a relatively low level of truncation.\cite{Li12266403}
We will show in \Sec{thnum} that these features will largely
remain in the evaluations of nonequilibrium properties.

\section{Nonequilibrium response theory}
\label{ththeo}

\subsection{Linearity of the hierarchical Liouville space}
\label{ththeo1}

 To highlight the linearity of HEOM, we arrange
the involving ADOs in a column vector,
denoted symbolically as
\be\label{bfrho}
  {\bm\rho}(t)\equiv \big\{\rho(t),\,\rho^{(1)}_{j}\!(t),\,
   \rho^{(2)}_{j_1\!j_2}\!(t),\, \cdots\,\big\}.
\ee
Thus, \Eq{HEOM} can be recast in a matrix-vector form (each element of
the vector in \Eq{bfrho} is a matrix) as follows,
\be
 \dot{\bm\rho}=-i\bfL(t)\bm\rho, \label{heom-mat-vec}
\ee
with the time-dependent hierarchical-space Liouvillian, as inferred
from \Eqs{HEOM}--(\ref{calLt}), being of
\be\label{dfLsum}
  \bfL(t)= \bfL_s+\delta\mathcal{L}(t){\bfone} + \delta{\bfV}(t)\, .
\ee
It consists not just the time-independent $\bfL_s$
part, but also two time-dependent parts and each of them will
be treated as perturbation at the linear response level soon.
Specifically, $\delta\mathcal{L}(t){\bfone}$,
with  ${\bfone}$ denoting the unit operator in the hierarchical Liouville space,
is attributed to a time-dependent external field acting on the reduced system,
while $\delta{\bfV}(t)$ is diagonal and due to the time-dependent potentials
$\delta\Delta_{\alpha}(t)$ applied to electrodes.

We may denote $\delta\Delta_{\alpha}(t)=x_{\alpha}\delta\Delta(t)$,
with $0\leq x_{\rm L}\equiv 1+x_{\rm R} \leq 1$; thus
$\delta\Delta(t)=\delta\Delta_{\rm L}(t)-\delta\Delta_{\rm R}(t)$. It
specifies the additional time-dependent bias voltage, on top of the
constant $V=\mu_{\rm L}-\mu_{\rm R}$, applied across the two
reservoirs. As inferred from \Eq{gammaSum}, we have then
\be\label{del_bfL_prime}
  \delta{\bfV}(t)=-{\bfS}\,\delta\Delta(t) ,
\ee
where ${\bfS}\equiv \text{diag}\big\{0,S^{(n)}_{j_1\cdots j_n};n=1,\cdots,L\big\}$,
with
\be\label{Sn}
  S^{(n)}_{j_1\cdots j_n}\equiv \sum_{r=1}^n
   \big(\sigma x_{\alpha}\big)_{\sigma,\alpha\,\in j_r}.
\ee
Note that  $S^{(0)}=0$.

The additivity of \Eq{dfLsum} and the linearity of HEOM lead readily to
the interaction picture of the HEOM dynamics in response to the
time-dependent external disturbance $\delta\bfL(t)=\delta{\mathcal
L}(t)\bfone +\delta{\bfV}(t)$. 
The initial unperturbed ADOs vector assumes the nonequilibrium
steady-state form of
\be\label{st_ADOs}
 {\bm\rho}^\text{st}(T,V)\equiv \big\{\bar\rho,\,
    \bar\rho^{(1)}_{j},\,
   \bar\rho^{(2)}_{j_1\!j_2},\,\cdots\,\big\},
\ee
under given temperature $T$ and constant bias voltage $V$.
It is obtained as the solutions to the linear equations,
$\bfL_s{\bm\rho}^\text{st}(T,V)=0$, subject to the normalization
condition for the reduced density
matrix.\cite{Jin08234703,Zhe08184112,Zhe09164708}
The unperturbed HEOM propagator reads $\bfG_s(t)\equiv
\exp(-i\bfL_st)$. Based on the first-order perturbation theory,
$\delta\bm\rho(t) \equiv \bm\rho(t)- {\bm\rho}^\text{st}(T,V)$ is then
\be\label{del_bfrho}
 \delta\bm\rho(t)=-i\int_{0}^t\! {d}\tau\,
  \bfG_s(t-\tau)\delta{\bfL}(\tau){\bm\rho}^\text{st}(T,V).
\ee

The response magnitude of a local system observable $\hat A$ is
evaluated by the variation in its expectation value, $\delta A(t) =
{\rm Tr}\{\hat A\delta\rho(t)\}$. Apparently, this involves the
zeroth-tier ADO $\delta\rho(t)$ in
$\delta\bm\rho(t)
 \equiv \big\{\delta\rho(t),\,
   \delta\rho^{(n)}_{j_1\cdots j_n}(t);  n=1,\cdots,L\big\}$.
In contrast, the response current under applied voltages cannot be
extracted from $\delta\rho(t)$, because the current operator is not a
local system observable. Instead, as inferred from \Eq{curr_t}, while
the steady-state current $\bar I_{\alpha}$ through $\alpha$-reservoir
is related to the steady-state first-tier ADOs, $\bar\rho^{(1)}_j\equiv
\bar\rho^{\sigma}_{\alpha\mu m}$, the response time-dependent current,
$\delta I_{\alpha}(t)=I_{\alpha}(t)-\bar I_{\alpha}$, is obtained via
$\delta\rho^{(1)}_j(t)=\delta \rho^{\sigma}_{\alpha\mu m}(t)$.

The above two situations will be treated respectively, by considering
$\delta{\bfL}(\tau)=\delta{\mathcal L}(t)\bfone$ and
$\delta{\bfL}(\tau)=\delta\bfV(t)$, in the coming two subsections:
\Sec{ththeo2} treats the local system response to a time-dependent
external field acting on the reduced system, while \Sec{ththeo3}
addresses the issue of electric current response to external voltage
applied to reservoirs.

\subsection{Nonequilibrium correlation and response functions of system}
\label{ththeo2}

Let $\hat A$ and $\hat B$ be two arbitrary local system operators, and
consider the correlation functions, $C_{AB}(t-\tau)=\la \hat A(t)\hat
B(\tau) \ra_{\rm st}$ and $S_{AB}(t-\tau)= \la \{\hat A(t),\hat
B(\tau)\} \ra_{\rm st}$, and response function, $\chi_{AB}(t-\tau)=i\la
[\hat A(t),\hat B(\tau)] \ra_{\rm st}$. It is well known that for the
equilibrium case they are related to each other via the
fluctuation-dissipation theorem. The nonequilibrium case is rather
complicated, and the relation between nonequilibrium correlation and
response functions is beyond the scope of the present paper.

We now focus on the evaluation of local system correlation/response
functions with the HEOM approach. This is based on the equivalence
between the HEOM-space linear response theory of \Eq{del_bfrho} and
that of the full system-plus-bath composite space.

We start with the evaluation of nonequilibrium steady-state correlation
function $C_{AB}(t)=\la \hat A(t)\hat B(0) \ra_{\rm st}$, as follows.
By definition, the system correlation function can be recast into the
form of
\begin{align}\label{Cab_def}
 C_{AB}(t)
&={\rm Tr}_{\T} \big\{\hat A{\cal G}_{\T}(t)
 [\hat B\rho^{\rm st}_{\T}(T,V)]\big\}
\nl&
\equiv {\rm Tr}_{\T}[\hat A \ti\rho_{\T}(t)]
\nl&
 ={\rm Tr}[\hat A \ti\rho(t)].
\end{align}
The $\rho^{\rm st}_{\T}(T,V)$ and ${\cal G}_{\T}(t)$ in the first
identity are the steady-state density operator and the propagator,
respectively, in the total system-bath composite space under constant
bias voltage $V$. Define in the last two identities of \Eq{Cab_def} are also
 $\ti\rho_{\T}(t)\equiv{\cal G}_{\T}(t)\ti\rho_{\T}(0)$
and $\ti\rho(t)\equiv {\rm tr}_{\B}\ti\rho_{\T}(t)$,
with
 $\ti\rho_{\T}(0) = \hat B\rho^{\rm st}_{\T}(T,V)$.
Equation (\ref{Cab_def}) can be considered in terms of the linear
response theory, in which the perturbation Liouvillian induced by an
external field $\delta\epsilon(t)$ assumes the form of $-i\delta{\cal
L}(t)(\cdot)=\hat B(\cdot)\delta\epsilon(t)$, followed by the
observation on the local system dynamical variable $\hat A$. Both $\hat
A$ and $\hat B$ can be non-Hermitian. Moreover, $\delta{\cal L}(t)$ is
treated formally as a perturbation and can be a one-side action rather
than having a commutator form.

For the evaluation of $C_{AB}(t)$ with the HEOM-space dynamics, the
corresponding perturbation Liouvillian is $\delta{\bfL}(t)=\delta{\cal
L}(t)\bfone$, with the above defined $\delta{\cal L}(t)$. It leads to
$-i\delta{\bfL}(\tau){\bm\rho}^{\rm st}(T,V) =\hat B {\bm\rho}^{\rm
st}(T,V) \delta\epsilon(\tau)$ involved in \Eq{del_bfrho}. The linear
response theory that leads to the last identity of \Eq{Cab_def} is now
of the $\ti\rho(t)$ being just the zeroth-tier component of
\be\label{tibmrho_t}
 \ti{\bm\rho}(t)\equiv \big\{\ti\rho(t),\,\ti\rho^{(1)}_{j}\!(t),\,
   \ti\rho^{(2)}_{j_1\!j_2}\!(t),\,\cdots\,\big\}
={\bfG}_s(t)\ti{\bm\rho}(0), \ee
with the initial value of [\emph{cf.}~\Eq{st_ADOs}]
\be\label{tibmrho_init1}
 \ti{\bm\rho}(0)=\hat B{\bm\rho}^{\rm st}(T,V)
 = \big\{\hat B\bar\rho,\hat B\bar\rho^{(1)}_{j}\!,\,
    \hat B\bar\rho^{(2)}_{j_1\!j_2},\,\cdots\,\big\}\,.
\ee
The HEOM evaluations of $S_{AB}(t)$ and $\chi_{AB}(t)$
are similar, but with the initial ADOs of
$\ti{\bm\rho}(0)=\{\hat B,{\bm\rho}^{\rm st}(T,V)\}$
and $i[\hat B,{\bm\rho}^{\rm st}(T,V)]$, respectively.

Care must be taken when propagating \Eq{tibmrho_t}, for the HEOM
propagator ${\bfG}_s(t)$ involving the Grassmann superoperators ${\cal
A}_{\bar j}$ and ${\cal C}_j$ defined in \Eq{calAC}. Note that the
steady-state system density operator $\bar\rho$ is always of the
Grassmann-even (or bosonic) parity. Therefore, the zeroth-tier ADO
$\ti\rho(t)$ in the above cases is of the same Grassmann parity as the
operator $\hat B$, while the ADOs at the adjacent neighboring tier
level are of opposite parity. The HEOM propagation in \Eq{tibmrho_t} is
then specified accordingly.

It is also worth pointing out that the HEOM evaluation of equilibrium
correlation and response functions of the local system can be
simplified when $J_{\rm L}(\omega)\propto J_{\rm R}(\omega)$. In this
case, two reservoirs can be combined as a whole entity bath, resulting
in a HEOM formalism that depends no longer on the reservoir-index
$\alpha$.

\subsection{Current response to applied bias voltages}
\label{ththeo3}

\subsubsection{Dynamic differential admittance}
\label{ththeo3.1}

Consider first the differential circuit current through a two-terminal
transport system composed of an quantum impurity and two leads, $\delta
I(t)=\frac{1}{2}[\delta I_{\rm L}(t) -\delta I_{\rm R}(t)]$, in
response to a perturbative shift of reservoir chemical potential
$\delta\Delta(t)$.

We have
\be\label{Galp_t}
 \delta I_{\alpha}(t)=\int_{0}^{t}\! {d}\tau\,
   G_{\alpha}(t-\tau)\,\delta\Delta(\tau).
\ee
The HEOM-space dynamics results in
\be\label{Galp_t_HEOM}
  G_{\alpha}(t)= 2\, {\rm Re} \sum_{\mu m}
   {\rm Tr}\left[\hat a^{+}_{\mu}
    \ti \rho^{-}_{\alpha\mu m}(t)\right],
\ee
with $\ti \rho^{-}_{\alpha\mu m}(t)$ denoting the first-tier ADOs in
$\ti{\bm\rho}(t)$ [\Eq{tibmrho_t}] with the initial value of
[\emph{cf.}\ \Eqs{del_bfL_prime}-(\ref{st_ADOs})]
\be\label{tibmrho_init2}
 \ti{\bm\rho}(0)= -{\bfS}{\bm\rho}^\text{st}\!(T,\!V)
\equiv -\big\{0,S^{(1)}_{j}\!\bar\rho^{(1)}_{j}\!,\!
          S^{(2)}_{j_1\!j_2}\bar\rho^{(2)}_{j_1\!j_2},\!\cdots\!
         \big\}.
\ee
Denote the half-Fourier transform,
\be\label{Galp_w}
  G_{\alpha}(\omega) \equiv \int_0^{\infty}\!\! {d}t\,
    e^{i\omega t}G_{\alpha}(t).
\ee
The admittance is given by $G(\omega)= \frac{1}{2}[G_{\rm L}(\omega)- G_{\rm
R}(\omega)]$, with its zero-frequency component recovering the
steady-state differential conductance as $d\bar I/dV = G(\omega=0)$.

\subsubsection{Current-number and current-current response functions}
\label{ththeo3.2}

Consider now the differential current $\delta I_{\alpha}(t)$ in
response to an additional time-dependent chemical potential
$\delta\Delta_{\alpha'}(t)$ applied on a specified $\alpha'$-reservoir.
Note that the Hamiltonian of the total composite system, \Eq{Htotal},
is now subject to a perturbation of $\delta H_{\T}(t)= \hat
N_{\alpha'}\delta\Delta_{\alpha'}(t)$, with $\hat N_{\alpha'}=\sum_k
\hat d^{\dg}_{\alpha' k}\hat d_{\alpha' k}$ being electron number
operator of the $\alpha'$-reservoir. Thus, the hierarchical Liouville
space linear response theory leads to
\be \label{delI_resp1}
 \delta I_{\alpha}(t)=\int_{0}^t\! {d}\tau\,
    G_{\alpha\alpha'}(t-\tau)\, \delta\Delta_{\alpha'}(\tau) \, ,
\ee
where the kernel is characterized by the nonequilibrium steady-state
current-number response function,
\be\label{g_alp_t}
  G_{\alpha\alpha'}(t-\tau) = -i\, \la [\hat I_{\alpha}(t), \hat N_{\alpha'}(\tau)]
   \ra_{\rm st} \, ,
\ee
with $\la (\cdot)\ra_{\rm st} \equiv {\rm Tr}_{\T}[(\cdot)\rho^{\rm
st}_{\rm T}(T,V)]$.
Equation~\eqref{g_alp_t} can be derived by following \Eqs{curr_t},
(\ref{del_bfrho}) and (\ref{delI_resp1}), and the HEOM evaluation of
$G_{\alpha\alpha'}(t-\tau)$ can be achieved as follows.
Equation~\eqref{del_bfL_prime} is recast as
$\delta{\bfL}(t)=-{\bfS}_{\alpha'}\delta\Delta_{\alpha'}(t)$, where
${\bfS}_{\alpha'} =\text{diag}\big\{0,
    S^{\alpha'}_{j_1\cdots j_n}\big\}$,
where $S^{\alpha'}_{j_1\cdots j_n}$ is similar to $S^{(n)}_{j_1\cdots
j_n}$ of \Eq{Sn} but with $x_{\alpha}=\delta_{\alpha\alpha'}$.
Therefore,
\be\label{Salpha}
 S^{\alpha'}_{j_1\cdots j_n} = \sum_{r=1}^n 
   (\sigma \delta_{\alpha\alpha'})_{\sigma\alpha \in j_r} \, .
\ee
Its rhs collects the signs ($\sigma = +1$ or $-1$) in the involving
$(j\equiv\{\sigma\alpha\mu m\})$-indexes whenever $\alpha=\alpha'$.
The suitable initial values for the vector of ADOs are
$$ 
 \ti{\bm\rho}_{\alpha'}(0)=-{\bfS}_{\alpha'}{\bm\rho}^\text{st}(T,V)
  =-\left\{0,S^{\alpha'}_{j}\!\bar\rho^{(1)}_{j}\!,
          S^{\alpha'}_{j_1\!j_2}\bar\rho^{(2)}_{j_1\!j_2},\!\cdots\!
         \right\},
$$ 
followed by the unperturbed HEOM-space evolution,
\be\label{rhot_alpha}
 \ti{\bm\rho}_{\alpha'}(t) = {\bfG}_s(t)\ti{\bm\rho}_{\alpha'}(0)
 \equiv \big\{\ti\rho(t;\alpha'),\ \ti\rho^{(1)}_{j}\!(t;\alpha'),\,\cdots \big\}.
\ee
The involving first-tier ADOs, $\ti\rho^{(1)}_j(t;\alpha')\equiv
\ti\rho^{\sigma}_{\alpha\mu m}(t;\alpha')$, are used to evaluate the
current-number response function [\emph{cf}.~\Eq{Galp_t_HEOM}]:
\be\label{gt_final}
  G_{\alpha\alpha'}(t) = 2\, {\rm Re} \sum_{\mu m}
   {\rm Tr}\left[\hat a^{+}_{\mu}
    \ti\rho^{-}_{\alpha\mu m}(t;\alpha')\right] .
\ee Apparently, $G_{\alpha}(t)= x_{\rm L} G_{\alpha{\rm L}}(t)+x_{\rm
R}G_{\alpha{\rm R}}(t)$, which is just the dynamic admittance
considered in \Sec{ththeo3.1}.

The nonequilibrium steady-state current-current response function,
$\chi_{\alpha\alpha'}(t)$, can be obtained numerically by taking the
time derivative of $G_{\alpha\alpha'}(t)$,
\be\label{chi_current}
 \chi_{\alpha\alpha'}(t)\equiv i \la [\hat I_{\alpha}(t),\hat I_{\alpha'}(0)]\ra_{\rm st}
  = {\dot G}_{\alpha\alpha'}(t).
\ee
In the hierarchical Liouville space, $\chi_{\alpha\alpha'}(t)$ can be
explicitly expressed by the zeroth-, first- and second-tier ADOs, as
inferred from \Eq{gt_final} and the EOM for $\ti{\rho}^{-}_{\alpha\mu
m}(t;\alpha')$. Its Fourier transform, the current-current response
spectrum, may carry certain information about the shot noise of the
impurity system.

In general, the correlation/response functions between an arbitrary
local system operator $\hat A$ and the electron number operator $\hat
N_{\alpha'}$ of the $\alpha'$-electrode can be evaluated via the
zeroth-tier ADO $\ti\rho(t;\alpha')$ of \Eq{rhot_alpha}, such as $i\la
[\hat A(t), \hat N_{\alpha'}(0)]\ra_{\rm st}=-{\rm Tr}[\hat A\,
\ti\rho(t;\alpha')]$, by using the HEOM Liouville propagator.
Its time derivative gives $i\la [\hat A(t), \hat
I_{\alpha'}(0)]\ra_{\rm st}$.

\section{Results and discussions}
\label{thnum}

We now demonstrate the numerical performance of the HEOM approach on
evaluation of nonequilibrium response properties of quantum impurity
systems. The hierarchical Liouville-space linear response theory
established in \Sec{ththeo} is employed to obtain the relevant
correlation/response functions, from which the response properties are
extracted.

It is worth emphasizing that the numerical examples presented in this
section aim at verifying the accuracy and universality of the proposed
methodology, rather than addressing concrete physical
problems. To this end, the widely studied standard single-impurity Anderson
model (SIAM) is considered.
The Hamiltonian of the single-impurity is
$H_\text{sys}=\epsilon_{\uparrow} \hat{n}_{\uparrow}+ \epsilon_{\downarrow}
\hat{n}_{\downarrow} + U\hat{n}_{\uparrow}\hat{n}_{\downarrow}$, with
$\hat{n}_{\mu}=\hat a^{\dg}_{\mu}\hat a_{\mu}$ being the electron
number operator for the spin-$\mu$ ($\uparrow$ or $\downarrow$)
impurity level. The impurity is coupled to two noninteracting electron
reservoirs ($\alpha = $ L and R). For simplicity, the spectral
(or hybridization) function of $\alpha$-reservoir assumes a diagonal
and Lorentzian form, \emph{i.e.},
$J_{\alpha\mu\nu}(\omega)=\delta_{\mu\nu}
\frac{\Gamma_{\alpha}W^{2}_{\alpha}}{2[(\omega-\mu_{\alpha})^{2}+W^{2}_{\alpha}]}$,
with $\Gamma_{\alpha}$ and $W_\alpha$ being the linewidth and bandwidth
parameters, respectively.

Note that the same set of system parameters are adopted for all
calculations (except for specially specified): $\epsilon =
\epsilon_{\uparrow} = \epsilon_{\downarrow} = -0.5$, $U=1.5$, $T=0.02$, $\Gamma =
\Gamma_{\rm L} = \Gamma_{\rm R} = 0.1$, $W_{\rm L} = W_{\rm R} = 2$,
all in units of meV.
The nonequilibrium situation concerns a steady state defined by a fixed
bias voltage applied antisymmetrically to the two reservoirs,
\emph{i.e.}, $\mu_{\rm L} = -\mu_{\rm R} = \frac{V_{0}}{2}$ with
$V_{0} = -V = 0.2$, and/or $0.7\,$meV.
A recently developed $[N\!-\!1/N]$ Pad\'{e} spectrum decomposition
scheme\cite{Hu10101106, Hu11244106} with $N=8$ (i.e., $M=9$) is used
for the efficient construction of the hierarchical Liouville propagator
associated with \Eq{HEOM}.

To obtain quantitatively converged numerical results, we increase the
truncation level $L$ and the number of exponential terms $M$
continually until convergence is reached.
Table~\ref{table1} lists the probabilities that the impurity is singly
occupied by spin-$\mu$ electron ($P_\mu = \la\mu|\bar\rho(T,V)|\mu\ra$
with $\mu = \uparrow$ or $\downarrow$); or doubly occupied ($P_{\uparrow\downarrow}
=\la {\uparrow\downarrow} | \bar\rho(T,V) | {\uparrow\downarrow} \ra$).
Here, $\bar\rho(T,V)$ is the nonequilibrium steady-state reduced density
matrix under temperature $T$ and antisymmetric applied voltage $V$.
Calculations are done at different truncation level $L$ (up to $L=5$)
and fixed $M=9$. Apparently, the HEOM results converge rapidly
and uniformly with the increasing $L$, \emph{i.e.}, with higher-tier
ADOs included explicitly in \Eq{HEOM}. In particular, the remaining
relative deviations between the results of $L=4$ and $L=5$ are less than
0.1\%, indicating that the $L=4$ level of truncation is sufficient for
the present set of parameters.
It is also affirmed $M=9$ is sufficient to yield convergent results;
see Supplemental Material.\cite{SM_skw}
These are further affirmed by the calculated steady-state current $\bar{I}(V)$
across the impurity, which also converges quantitatively with rather minor
residual uncertainty at $L = 4$ and $M = 9$.
Note also that the truncation at $L=1$ level results in the sequential
current contribution, which is negligibly small for the present
nonequilibrium system setup. The values of $\bar I$ evaluated at different
truncation levels clearly indicate the current contributions from different
orders of cotunneling processes.

\begin{table}{}
\begin{tabular}{c|c|c|c}
\hline \hline
 $L$ & $P_{\mu}$  & $P_{\uparrow\downarrow}$ & $\ \bar I$ (nA) \\
\hline
1  &  \, 0.500 ({\it 0.500}) &  0.001 ({\it 0.000}) &  0.003 \\
\hline
2  &  \, 0.441 ({\it 0.462}) &  0.025 ({\it 0.027}) &  4.654 \\
\hline
3  &  \, 0.439 ({\it 0.454}) &  0.024 ({\it 0.025}) &  4.920 \\
\hline
4  &  \, 0.440 ({\it 0.457}) &  0.024 ({\it 0.024}) &  4.799 \\
\hline
5  &  \, 0.440 ({\it 0.457}) &  0.024 ({\it 0.024}) &  4.799 \\
\hline \hline
\end{tabular}
\caption{ Spin-$\mu$ single- and double-occupation probabilities
($P_{\mu}$ and $P_{\uparrow\downarrow}$), and steady-state current of
an SIAM with two electrons reservoirs under an antisymmetrically
applied bias voltage of $V_{0} = -V = 0.7\,$meV. Calculations are done
by solving the HEOM of \Eq{HEOM} truncated at different level $L$.
The parameters are adopted are (in units of meV):
$\epsilon=\epsilon_{\uparrow}=\epsilon_{\downarrow}=-0.5$, $U=1.5$,
$\Gamma_{\rm L}=\Gamma_{\rm R}=0.1$, $W_{\rm L}=W_{\rm R}=2$, and
$T=0.02$. For comparison, the numbers of equilibrium situation of $V_{0} =
0$ are shown in the parentheses.}
\label{table1}
\end{table}

\begin{figure}
\includegraphics[width=0.9\columnwidth]{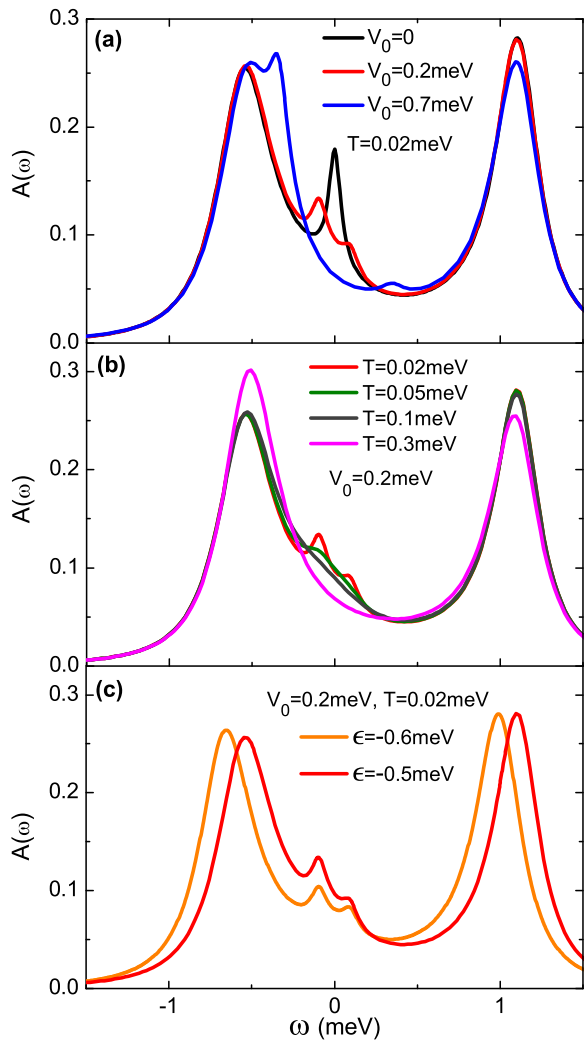}
\caption{The HEOM calculated spectral function of an SIAM system, $A(\omega)=A_{\uparrow}(\omega)
=A_{\downarrow}(\omega)$, in unit of $(\pi\Gamma)^{-1}$.
The parameters adopted are specified in the caption of Table~\ref{table1}.
The three panels exhibit the variations of $A(\w)$, particularly the
evolution of the Kondo and Hubbard peaks, versus
(a) the applied bias voltage $V_{0}$,
(b) the temperature $T$, and
(c) the shift of impurity level energy $\epsilon$ by a gate voltage, respectively.
} \label{fig1}
\end{figure}

\begin{figure}
\includegraphics[width=0.9\columnwidth]{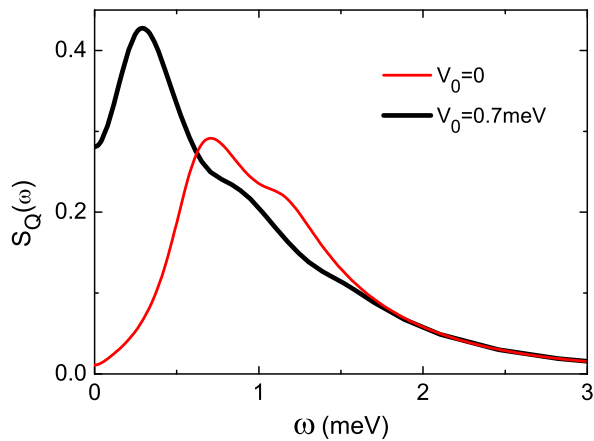}
\caption{HEOM calculated local charge fluctuation spectrum,
$S_{Q}(\w)$ of \Eq{def-sqq}, in unit of $e^2$.
The parameters adopted are the same as those specified
in caption of Table~\ref{table1}.}
\label{fig2}
\end{figure}

\begin{figure}
\includegraphics[width=0.9\columnwidth]{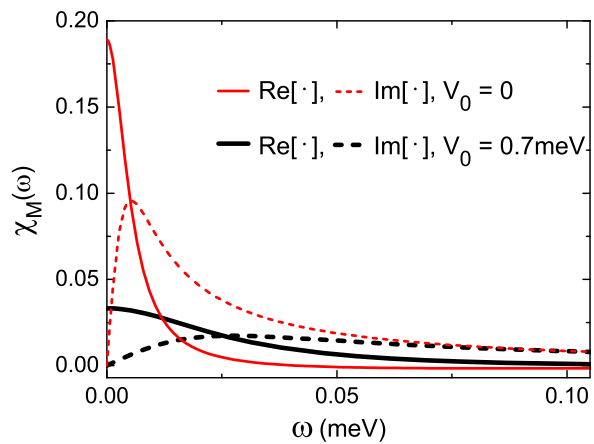}
\caption{HEOM calculated local magnetic susceptibility,
$\chi_{M}(\omega)$ of \Eq{def-chi}, in unit of $g^{2}\mu^{2}_{B}/k_{B}T$.
The parameters adopted are the same as those specified
in caption of Table~\ref{table1}.
}
\label{fig3}
\end{figure}

In the following, we first show the spectral function of the SIAM
calculated by using the HEOM approach (see \Fig{fig1}), and then
present the evaluation of some typical response properties in both
equilibrium and nonequilibrium situations. These will include the
local charge fluctuation spectrum $S_{Q}(\omega)$ (see \Fig{fig2}), local
magnetic susceptibility $\chi_{M}(\omega)$ (see \Fig{fig3}), and differential
admittance $G(\omega)$ (see \Fig{fig5}).
\emph{All calculations are carried out
at the truncation level of $L = 4$ and $M=9$.} Based on the analysis
of Table~\ref{table1}, the resulting response properties are expected
to be quantitatively converged with respect to $L=4$ and $M=9$.

Figure \ref{fig1} depicts the HEOM calculated spin-$\mu$ spectral
function of the impurity,
\be\label{Aomega}
 A_{\mu}(\omega)=\frac{1}{\pi}\, {\rm Re} \left\{
 \int_{0}^{\infty}\!\! d t\, e^{i\omega t} \left\la
 \left\{\hat a_{\mu}(t), \hat a^{\dg}_{\mu}(0) \right\}
 \right\ra_{\rm st}\, \right\}.
\ee
The effect of bias voltage $V_{0}$ on $A_\mu(\omega)$ is illustrated in
\Fig{fig1}(a). Clearly, the equilibrium $A_\mu(\omega)$ reproduces correctly the
well known features of SIAM, such as the Hubbard peaks at around $\omega =
\epsilon$ and $\omega = \epsilon + U$, and the Kondo peak centered at $\omega =
\mu^{\rm eq} = 0$.\cite{Hew93}
In Ref.\,\onlinecite{Li12266403}, the equilibrium $A_\mu(\omega)$ of
SIAM in the Kondo regime has been investigated with the HEOM approach
thoroughly and the existence of Kondo resonance is manifested by the
correct universal scaling behavior there.
In the nonequilibrium situation where an external voltage is applied
antisymmetrically to the L and R reservoirs, the Hubbard peaks remain
largely unchanged in both position and height.
In contrast, the Kondo peak is split by the voltage into two, which
appear at $\omega =\frac{V_{0}}{2}$ and $\omega = -\frac{V_{0}}{2}$ and
correspond to the shifted reservoir chemical potentials $\mu_{\rm L}$
and $\mu_{\rm R}$, respectively.
Obviously, as the bias voltage $V_{0}$ increases from $0$ to $0.2\,$meV,
then to $0.7\,$meV, the progressive splitting of Kondo peak is observed
in \Fig{fig1}(a).
Figure~\ref{fig1}(b) plots the calculated $A(\omega)$ of the same SIAM
system at various temperatures. Apparently, as the temperature
increases over an order of magnitude, the two Hubbard resonance peaks
at $\omega=\epsilon$ and $\omega=\epsilon+U$ almost remain intact. In
contrast, the split peaks at $\omega=\mu_{L}$ and $\mu_{R}$ vanish
quickly at the higher temperature. This clearly highlights the strong
electron correlation features in the present nonequilibrium SIAM.
To further verify that the split peaks near $\w = 0$ are of Kondo
nature, we examine the variation of $A(\w)$ versus a gate voltage
applied to the dot. The gate voltage is considered to shift the
impurity level energy $\epsilon$ by $0.1\,$meV, and the corresponding
change of calculated $A(\w)$ is shown in \Fig{fig1}(c). Apparently, as
$\epsilon$ drops from $-0.5$ to $-0.6\,$meV by the gate voltage, the
two Hubbard resonance peaks at $\omega=\epsilon$ and
$\omega=\epsilon+U$ are displaced by $0.1\,$meV. In contrast, the two
peaks at $\w = \pm \frac{V_0}{2}$ remain pinned to the reservoir
chemical potentials $\mu_{L}$ and $\mu_{R}$, indicating that these two
peaks are indeed of Kondo origin.

Usually, the Hubbard peaks at $\omega = \epsilon$ and $\omega =
\epsilon + U$ converge more rapidly than the resonance peaks at
$\omega=\pm\frac{V_{0}}{2}$ when truncation level $L$ increases. This
also reflects the Kondo nature of resonance peaks at $\omega
=\pm\frac{V_{0}}{2}$.\cite{SM_skw} Moreover, there is a long-time
oscillatory tail in the real time evolution which is crucial for the
appearance of Kondo peaks. We also find that the short-time dynamics of
the retarded Green's function and high-frequency part of $A(\omega)$
converge more rapidly when the truncation level $L$ increases. In other
words, in the HEOM framework, one can extract the spectral function at
high frequency range at a relatively lower truncation level and
relatively shorter evolution time than those at resonance frequencies,
without compromising the accuracy.

We then exemplify the numerical tractability of HEOM approach via
evaluation of three types of response properties. These include the
local charge fluctuation spectrum $S_{Q}(\omega)$, local magnetic
susceptibility $\chi_M(\omega)$, and differential admittance spectrum
$G_{\alpha\alpha'}(\omega)$, defined respectively as follows,
\begin{align}
 S_{Q}(\omega) &\equiv \int^{\infty}_{-\infty} dt\, e^{i\omega t}
  \big\langle \big\{\Delta \hat{Q}(t),\Delta \hat{Q}(0)\big\}
  \big\rangle_{\rm st}\,,
\label{def-sqq} \\
  \chi_{M}(\omega) &\equiv i\int_{0}^{\infty}  dt\,
   e^{i\omega t}\, \big\la \big[\hat M(t), \hat M(0)
   \big]\big\ra_{\rm st}\,,
\label{def-chi} \\
  G_{\alpha\alpha'}(\omega) &\equiv -i \int_{0}^\infty
  dt\, e^{i\omega t}\, \big\la \big[\hat I_{\alpha}(t),
  \hat N_{\alpha'}(0) \big] \big\ra_{\rm st}\,.
\label{def-g}
\end{align}
In \Eq{def-sqq}, $\Delta \hat{Q}(t)=\hat{Q}(t)-\langle
\hat{Q}\rangle_{\rm st}$, with $\hat{Q}= \sum_\mu \hat{n}_\mu$ being
the total impurity occupation number operator. Therefore, $\Delta
\hat{Q}(t)$ describes the fluctuation of occupation number around the
averaged value.
For $\chi_M(\omega)$ of \Eq{def-chi}, $\hat M = g\mu_{B}\hat{S}_z$ is the
impurity magnetization operator, which originates from the electron
spin polarization induced by external magnetic field. Here, $g$ is the
electron $g$-factor, $\mu_B$ is the Bohr magneton, and $\hat{S}_z =
(\hat{n}_{\uparrow}-\hat{n}_{\downarrow})/2$ is the impurity spin
polarization operator.
In \Eq{def-g}, $G_{\alpha\alpha'}(\omega)$ is just the half-Fourier
transform of current-number response function of \Eq{g_alp_t}
or (\ref{gt_final}).
The time $t = 0$ in the individual \Eqs{def-sqq}--\eqref{def-g} represents the instant
at which the external perturbation (magnetic field or bias voltage) is
interrogated.

It is worth pointing out that all the three types of response
properties satisfy the following symmetry: the real (imaginary) part is
an even (odd) function of $\omega$.
This is due to the time-reversal symmetry of steady-state correlation
functions, \emph{i.e.}, $C_{AB}(t) = [C_{BA}(-t)]^\ast$.
In particular, for $S_{Q}(\omega)$ of \Eq{def-sqq}, $\hat{A} = \hat{B} =
\Delta\hat{Q}$. Consequently, it can be shown that $S_{Q}(\omega)$ is a
real function.
For clarity, Figs.~\ref{fig2}, \ref{fig3}, and \ref{fig5} will only
exhibit the $\omega \geq 0$ part of the dynamic response properties, while
the $\omega < 0$ part can be retrieved easily by applying the symmetry
relation.

The local charge fluctuation spectrum $S_{Q}(\omega)$ has been studied in
the context of shot noise of quantum dot systems.\cite{Agu04206601,
Fli05411, Luo07085325}
Figure~\ref{fig2} depicts the HEOM calculated $S_{Q}(\omega)$ of the SIAM
under our investigation. At equilibrium, the spectrum exhibits a
crossover behavior, where the two peaks centered at around $\omega =
|\epsilon| = 0.5$\,meV and $\omega = |\epsilon + U| = 1$\,meV largely overlap
and form a broad peak.
The positions of these two peaks correspond to the excitation energies
associated with change of impurity occupancy state.
In nonequilibrium situation, the crossover peak is observed to move to
a lower energy, since the chemical potential of reservoir R is drawn
closer to the impurity state by the applied voltage.

The local magnetic susceptibility $\chi_M(\omega)$ is a response property
of fundamental significance, particularly for strongly correlated
quantum impurity systems. It has been studied by various methods such
as NRG.\cite{Mer12075153}
Figure~\ref{fig3} displays the HEOM calculated $\chi_M(\omega)$ of the SIAM
of our concern.
In both equilibrium and nonequilibrium situations, the main
characteristic features of $\chi_M(\omega)$ appear at around zero energy.
Moreover, the magnitude of $\chi_M(\omega)$ is found to reduce
significantly in presence of applied bias voltage, especially in the
low energy range. This is consistent with the diminishing spectral
density at around zero-frequency due to the voltage-induced splitting
of Kondo peak; see \Fig{fig1}.

\begin{figure}
\includegraphics[width=0.9\columnwidth]{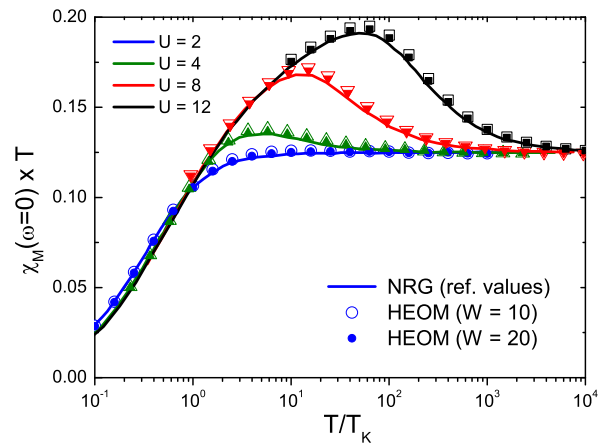}
\caption{Static local magnetic susceptibility multiplied by temperature,
$\chi_{M}(\omega = 0)\,T$ (in unit of $g^{2}\mu^{2}_{B}/k_{B}$),
versus $T/T_K$ for a series of equilibrium symmetric SIAM systems of different $U$.
Here, $T_K$ is the Kondo temperature, and $U$, $T$, and $W$ are in
unit of $\Gamma$. The HEOM results (scattered symbols) are compared with
the latest full density matrix NRG calculations (lines) of Ref.~\onlinecite{Mer12075153},
the Fig.~6 there. The NRG calculations are for very large reservoir bandwidth $W$,
while the HEOM results are obtained with relatively smaller bandwidths
($W = 10$ and $W = 20$) for saving computational cost.
}
\label{fig4}
\end{figure}

To verify the numerical accuracy of our calculated local magnetic
susceptibility, we compare the HEOM approach with the latest high-level
NRG method. The comparison is shown in \Fig{fig4}, where the equilibrium
static magnetic susceptibilities, $\chi_M(\omega = 0)$, of various symmetric
SIAM systems are calculated to reproduce the Fig.~6 of Ref.~\onlinecite{Mer12075153}.
Apparently, the HEOM and NRG results agree quantitatively with each other.

\begin{figure}
\includegraphics[width=0.9\columnwidth]{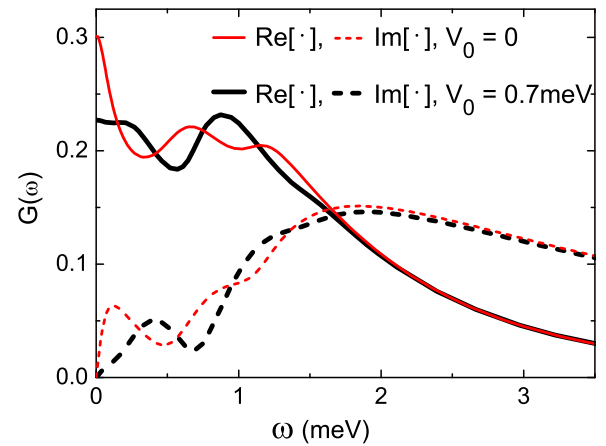}
\caption{HEOM calculated dynamic admittance,
$G(\omega) = \frac{1}{4}[G_{\rm LL}(\omega) + G_{\rm LR}(\omega) -
G_{\rm RL}(\omega) - G_{\rm RR}(\omega)]$ with $G_{\alpha\alpha'}(\omega)$ defined by
\Eq{def-g}, in unit of $e^2/h$.
The parameters adopted are the same as those specified
in caption of Table~\ref{table1}.
}
\label{fig5}
\end{figure}

The dynamic admittance is one of the most extensively studied response
properties of quantum dot systems. The frequency-dependent admittance
has been studied by scattering theory\cite{But93364, But934114,
Pre968130, But964793} and nonequilibrium Green's function
method.\cite{Win938487, Jau945528, Ana957632, Nig06206804, Wan07155336}
The HEOM approach has also been used to calculate the dynamic
admittance of noninteracting\cite{Zhe08184112} and interacting quantum
dots.\cite{Zhe08093016,Zhe09124508}
This was realized by calculating the transient current in response to a
delta-pulse voltage.\cite{Zhe08184112}
In the following, we revisit the evaluation for dynamic admittance
$G(\omega)$ via an alternative route, \emph{i.e.}, by calculating the
current-number response functions of \Eq{g_alp_t}.

Figure \ref{fig5} depicts the HEOM calculated differential admittance
of the SIAM under study, $G(\omega) = \frac{1}{4}[G_{\rm LL}(\omega) + G_{\rm LR}(\omega) -
G_{\rm RL}(\omega) - G_{\rm RR}(\omega)]=\frac{1}{2}[G_{L}(\omega)-G_{R}(\omega)]$;
\emph{cf.} \Eq{Galp_t}, with $G_{\alpha\alpha'}(\omega)$ defined
by \Eq{def-g}. Here we have chosen antisymmetrically applied probe ac bias,
$\delta\Delta_{L}(t)=-\delta\Delta_{R}(t)=\frac{1}{2}\delta\Delta(t)$.
As discussed extensively in Ref.\,\onlinecite{Zhe09164708},
the characteristic features of $G(\omega)$ appearing at around $\omega = |\epsilon|$
and $\omega = |\epsilon + U|$ correspond to the transitions between Fock states
of different occupancy, while the low-frequency features highlight the presence of
dynamic Kondo transition. Apparently, the dynamic Kondo transition is suppressed
by the applied voltage, which is analogous to the scenario of $\chi_M(\omega)$ as shown in
\Fig{fig3}.

\begin{figure}
\includegraphics[width=0.9\columnwidth]{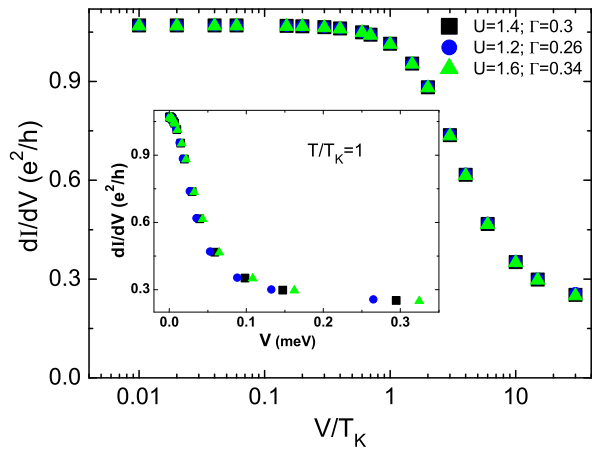}
\caption{
Calculated differential conductance $dI/dV$ of various symmetric SIAMs as a function of scaled voltage $V/T_{K}$.
Here, $T_K$ is the Kondo temperature calculated by $T_{K}=\frac{1}{2}\sqrt{\Gamma U}\, {\rm exp}(-\pi U/4\Gamma + \pi\Gamma/4U)$, with $\Gamma =\Gamma_{L}+\Gamma_{R}$.
Systems of three different combinations of parameters $U$ and $\Gamma$ (in units of meV) are demonstrated, with $T/T_{K}=1$ and $\Gamma_{L}=\Gamma_{R}$. The inset depicts $dI/dV$ versus unscaled voltage. The bandwidths of electrodes are $W_{\rm L}=W_{\rm R}=10\,$meV.
}
\label{fig6}
\end{figure}

At last we investigate the universal scaling properties of
nonequilibrium differential conductance $dI/dV$ (or the
zero-frequency admittance).
The universal scaling relation of nonequilibrium properties
of impurity systems with Kondo correlations have been studied.
\cite{Kam008154,Ros01156802,Pus04R513} For instance, Rosch~\emph{et al.}
have concluded that the nonequilibrium conductance at high voltages
scales universally with $V/T_K$; see the inset of Fig.~1 in
Ref.~\onlinecite{Ros01156802}.
To demonstrate the high accuracy of our HEOM approach in regimes
far from equilibrium, we calculate the conductance of various
symmetric SIAMs versus scaled and unscaled voltages, as displayed
in \Fig{fig6}.
Clearly, while the difference in $dI/dV$--$V$ becomes more accentuated
at a larger $V$ (see the inset of \Fig{fig6}), the $dI/dV$--$V/T_K$
exhibits a universal scaling relation for systems of different parameters.
Such a universal relation holds for all voltages examined (up to $V/T_K = 30$).
Therefore, the HEOM approach reproduces quantitatively the previously
predicted universal scaling relation for nonequilibrium conductance at
high voltages.

\section{Concluding remarks} \label{summary}

In this work, we develop a hierarchical dynamics approach for evaluation
of nonequilibrium dynamic response properties of quantum impurity systems.
It is based on a hierarchical equations of motion formalism, in conjunction
with a linear response theory established in the hierarchical Liouville space.
It provides a unified approach for arbitrary response and correlation functions
of local impurity systems, and transport current related response properties.

The proposed hierarchical Liouville-space approach resolves \emph{nonperturbatively}
the combined effects of \emph{e-e} interactions, impurity-reservoir dissipation,
and non-Markovian memory of reservoirs.
It provides a unified formalism for equilibrium and nonequilibrium dynamic
response properties of quantum impurity systems and can be applied to more
complex quantum impurity systems without extra derivation efforts.
Moreover, the HEOM results converge rapidly and uniformly with
higher-tier ADOs included explicitly and one can often obtain
convergent results at a relative low truncation level $L$.
With our present code and the computational resources at our disposal,
the lowest temperature that can be quantitatively accessed is $T \simeq
0.1\, T_K$ for a symmetric SIAM.

For equilibrium properties, our HEOM approach has achieved the same level
of accuracy as the latest state-of-the-art NRG method.\cite{Li12266403}
In this work, the accuracy of HEOM approach for calculations of nonequilibrium
properties is validated by reproducing some known numerical results or analytic
relations,\cite{Mer12075153,Ros01156802,Mei932601,Kon02125304} such as the static
local magnetic susceptibility, and the universal scaling relation of high-voltage
conductance.

In conclusion, the developed hierarchical Liouville-space approach
provides an accurate and universal tool for investigation of general
dynamic response properties of quantum impurity systems. In particular,
it addresses the nonequilibrium situations and resolves the full
frequency dependence details accurately.
It is thus potentially useful for exploration of quantum impurity systems and
strongly correlated lattice systems (combined with dynamical mean field theory),
which are of experimental significance for the advancement of nanoelectronics,
spintronics, and quantum information and computation.

\acknowledgments

The support from the Hong Kong UGC (AoE/P-04/08-2) and RGC (Grant
No.\,605012), the NSF of China (Grant No.\,21033008, No.\,21103157,
No.\,21233007, No.\,10904029, No.\,10905014, No.\,11274085), the
Fundamental Research Funds for Central Universities (Grant No.\,2340000034
and No.\,2340000025) (XZ), the Strategic Priority Research Program (B) of the CAS (XDB01020000) (XZ and YJY), and the Program for Excellent Young Teachers
in Hangzhou Normal University (JJ) is gratefully acknowledged.

\appendix

\section{}

In this Supplemental Material, we will discuss some details of the
numerical implementation of the proposed HEOM formalism to the
nonequilibrium dynamical properties of quantum impurity systems.

\begin{table}{}
\begin{tabular}{c|c|c|c}
\hline \hline
 $L, M$ & $P_{\mu}$  & $P_{\uparrow\downarrow}$ & $\ \bar I$ (nA) \\
\hline
1, 9  &  \, 0.500 ({\it 0.500}) &  0.001 ({\it 0.000}) &  0.003 \\
\hline
2, 9  &  \, 0.441 ({\it 0.462}) &  0.025 ({\it 0.027}) &  4.654 \\
\hline
3, 9  &  \, 0.439 ({\it 0.454}) &  0.024 ({\it 0.025}) &  4.920 \\
\hline
4, 9  &  \, 0.440 ({\it 0.457}) &  0.024 ({\it 0.024}) &  4.799 \\
\hline
4, 10  &  \, 0.440 ({\it 0.457}) &  0.0237 ({\it 0.024}) &  4.805 \\
\hline
4, 11  &  \, 0.440 ({\it 0.457}) &  0.0237 ({\it 0.0238}) &  4.806 \\
\hline
5, 9  &  \, 0.440 ({\it 0.457}) &  0.024 ({\it 0.024}) &  4.799 \\
\hline
5, 10  &  \, 0.440 ({\it 0.457}) &  0.0237 ({\it 0.024}) &  4.805 \\
\hline
5, 11  &  \, 0.440 ({\it 0.457}) &  0.0237 ({\it 0.0238}) &  4.806 \\
\hline
\hline
\end{tabular}
\caption{ Spin-$\mu$ single- and double-occupation probabilities
($P_{\mu}$ and $P_{\uparrow\downarrow}$), and steady-state current of
an SIAM with two electrons reservoirs under an antisymmetrically
applied bias voltage of $V_{0} = -V = 0.7\,$meV. Calculations are done
by solving the HEOM truncated at different level $L$ with different number
of exponential terms $M$. The parameters are the same as those specified
in caption of Table I of the main text. For comparison, the numbers of
equilibrium situation of $V_{0} = 0$ are shown in the parentheses.}
\label{table_SM_1}
\end{table}

\begin{figure}
\includegraphics[width=0.82\columnwidth]{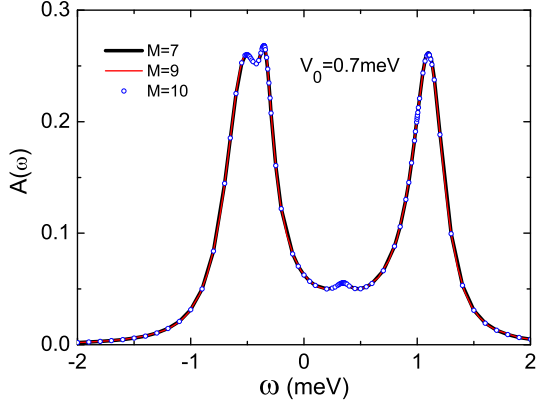}
\caption{
The convergence behavior of HEOM calculated spectral function $A(\omega)$ of SIAM
(in unit of $1/\pi\Gamma$) with respect to the number of exponential terms $M$
in spectrum decomposition. Here the truncation level is fixed at $L=4$.
The parameters are the same as those specified in caption of Table I of
the main text.
}
\label{figSM1}
\end{figure}

(i) In the main text, the spectral function of $\alpha$-reservoir assumes a diagonal
and Lorentzian form, \emph{i.e.}, $J_{\alpha\mu\nu}(\omega)=\delta_{\mu\nu}
\frac{\Gamma_{\alpha}W^{2}_{\alpha}}{(\omega-\mu_{\alpha})^{2}+W^{2}_{\alpha}}$, with
$\Gamma_{\alpha}$ and $W_\alpha$ being the linewidth and bandwidth parameters, respectively.
Then in the contour integration evaluation of $\tilde{C}^{\sigma;{\rm st}}_{\alpha\mu\nu}(t)
=\int_{-\infty}^{\infty}\! d\omega \frac{ e^{\sigma i\omega t}J^{\sigma}_{\alpha\mu\nu}(\omega
-\mu_{\alpha})}{1+e^{\sigma\beta(\omega-\mu_{\alpha})}}$, a total number of $M=N+1$ poles will
be involved, including one Drude pole of the reservoir spectral density and $N$ poles of the
Fermi distribution function.

In general, the number $K$ of distinct ADO indexes in the HEOM proposed in the main text
amounts to $K=2N_{\alpha}N_{\mu}M$, with $N_{\mu}$ being the number spin-orbitals of system
in direct contact to leads. The factor 2 accounts for the two choices of the sign variable
$\sigma$, while $N_{\alpha}=2$ for the distinct  $\alpha=$ L and R leads.

We have known the fact that overall computational cost increases dramatically
with both $L$ and $K$. To minimize the computational cost while maintaining the
quantitative accuracy, a highly efficient reservoir spectrum decomposition scheme
is needed to optimize the size of $K$. Various decomposition schemes have been
developed, including Matsubara spectrum decomposition,\cite{Jin08234703} partial
fractional decomposition, \cite{Cro09073102} Pad\'{e} spectrum decomposition (PSD),
\cite{Hu10101106, Hu11244106} and hybrid scheme.\cite{Zhe09164708}
To our knowledge, PSD scheme has the best performance and acquires the same
accuracy with the minimal size of $K$ until now. In the present work, the
$[N\!-\!1/N]$ Pad\'{e} spectrum decomposition scheme\cite{Hu10101106, Hu11244106}
is used for the efficient construction of the hierarchical Liouville propagator.

There is an overlap between Fermi distribution function $f^{(\sigma)}_{\alpha}
(\omega-\mu_{\alpha})$ and bath spectral density $J^{\sigma}_{\alpha\mu\nu}
(\omega-\mu_{\alpha})$ in the definition of bath correlation function
$\tilde{C}^{\sigma;{\rm st}}_{\alpha\mu\nu}(t)=\int_{-\infty}^{\infty}\! d\omega
e^{\sigma i\omega t}f^{(\sigma)}_{\alpha}(\omega-\mu_{\alpha})J^{\sigma}_{\alpha\mu\nu}
(\omega-\mu_{\alpha})$. Due to the finite bandwidth of reservoirs in realistic
quantum impurity system, the residual between $f^{(\sigma)}_{\alpha}(\omega-\mu_{\alpha})$
and its Pad\'{e} series expansion (with large enough $N$) at $\omega \gg W_{\alpha}$
will not introduce any approximation. Therefore, the exponential series expansion of
bath correlation function with large enough $M$ can efficiently resolves the non-Markovian
memory of reservoirs in both equilibrium and nonequilibrium situations.

To obtain quantitatively accurate numerical results, we increase the number of exponential
terms $M$ in spectrum decomposition continually until convergent results are arrived in
practice.\cite{Li12266403}

Table~\ref{table_SM_1} lists the probabilities that the impurity is singly
occupied by spin-$\mu$ electron ($P_\mu = \la\mu|\bar\rho(T,V)|\mu\ra$
with $\mu = \uparrow$ or $\downarrow$); or doubly occupied ($P_{\uparrow\downarrow}
=\la {\uparrow\downarrow} | \bar\rho(T,V) | {\uparrow\downarrow} \ra$).
Here, $\bar\rho(T,V)$ is the nonequilibrium steady-state reduced density
matrix under temperature $T$ and antisymmetric applied voltage $V$.
Calculations are done at different truncation level $L$ and different
number of exponential terms $M$ (up to $L=5$ and $M=11$). Apparently,
the HEOM results converge rapidly with the increasing $L$, \emph{i.e.},
with higher-tier ADOs included explicitly in HEOM. In particular,
the remaining relative deviations between the results of $L=4$ and $L=5$ are
less than 0.1\%, indicating that the $L=4$ level of truncation is sufficient
for the present set of parameters. At the same time, one can observe that
$M=9$ is sufficient to yield convergent results here.
These are further affirmed by the calculated steady-state current $\bar{I}(V)$
across the impurity, which also converges quantitatively with rather minor
residual uncertainty at $L = 4$ and $M=9$.

\Fig{figSM1} further illustrates the convergence behavior of $A(\omega)$ of SIAM with
respect to the number of exponential terms $M$ at the fixed truncation level $L=4$.
Obviously $M=9$ is large enough for the adopted parameters of the numerical examples in
the main text.

\begin{figure}
\includegraphics[width=0.82\columnwidth]{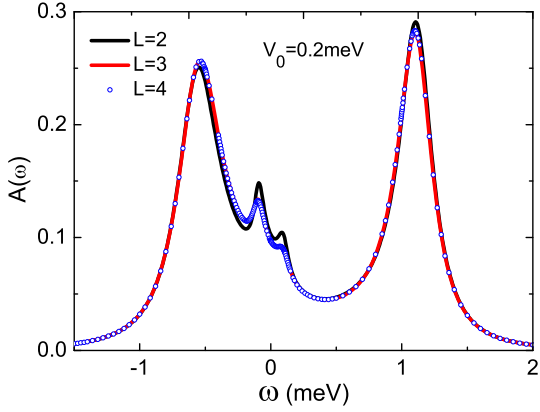}
\caption{
The convergence behavior of HEOM calculated spectral function $A(\omega)$ of SIAM
(in unit of $1/\pi\Gamma$) with respect to the truncation level $L=2,3,4$
with the fixed number of exponential terms $M=9$. The parameters are the
same as those specified in caption of Table I of the main text.
}
\label{figSM2}
\end{figure}

(ii) The terminal truncation level $L_{max}$ is often too high to reach
in practical implementation, and a truncation at a relatively low level
$L$ is inevitable. Here we adopt a straightforward truncation scheme, i.e.,
set all the higher-tier ADOs ($n > L$) zero directly. And we increase the
truncation level $L$ continually until the convergent numerical results
are obtained. For quantum impurity systems with nonzero e-e interactions,
calculations often converge rapidly and uniformly with the increasing truncation
level $L$. For weak to medium system-reservoir coupling strength, one can often
obtain quantitatively accurate results at a relatively low $L$.
HEOM calculated spectral function $A(\omega)$ of SIAM at different truncation level
$L=2$, $3$, $4$ with the fixed number of exponential terms in spectrum
decomposition ($M=9$) are plotted together in \Fig{figSM2}, to illustrate
the convergence behavior and gain the feeling on the numerical aspects of the
proposed formalism. Obviously, the numerical results converge rapidly and
uniformly with the increasing truncation level $L$.
\Fig{figSM2} further confirms the fact that $L=4$ level of truncation is
sufficient for the set of parameters adopted in the main text.
One can also observe that the resonance Hubbard peaks at around $\omega =
\epsilon$ and $\omega = \epsilon + U$ converge more easily than the split
Kondo peaks at $\omega =\pm\frac{V_{0}}{2}$ with increasing truncation level
$L$. This further confirms the strong correlation nature of resonance at
$\omega =\pm\frac{V_{0}}{2}$. Moreover, the numerical results at high and
far-from-resonance frequency range converge more easily than those at Kondo
resonance frequencies.

\begin{figure}
\includegraphics[width=0.82\columnwidth]{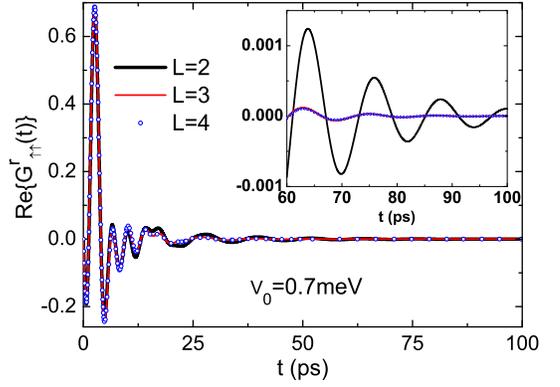}
\caption{
The convergence behavior of HEOM calculated $G^{r}_{\uparrow \uparrow}(t>0)$ of SIAM
with respect to the truncation level $L=2,3,4$, with fixed $M=9$. The parameters
are the same as those specified in caption of Table I of the main text. The inset
magnifies the long time oscillatory tail in real time evolution.
}
\label{figSM3}
\end{figure}

(iii) The hierarchical Liouville-space linear response theory proposed in
the main text can be implemented in either time-domain or frequency domain.
In the time-domain scheme, one first carry out the real-time evolution
under the unperturbed HEOM propagator ${\bfG}_s(t)$, starting from the
initial condition $\ti{\bm\rho}(0)$, i.e.,
\begin{equation}
\ti{\bm\rho}(t)={\bfG}_s(t)\ti{\bm\rho}(0).
\end{equation}
Then the correlation/response functions can be extracted from certain ADOs
components $\{\ti\rho^{(n)}_{j_{1}...j_{n}}\!(t)\}$ in $\ti{\bm\rho}(t)$,
as in Eq.\ (20) of the main text.  At last the frequency-dependent dynamical
properties are obtained straightforwardly by a half Fourier transform.

\Fig{figSM3} depicts the HEOM calculated real part of $G^{r}_{\uparrow \uparrow}(t>0)
=-\frac{i}{\hbar}\langle\{\hat{a}_{\uparrow}(t), \hat{a}^{\dg}_{\uparrow}(0)\}
\rangle_{\rm st}$ of SIAM system at different truncation level $L=2$, $3$, $4$.
The number of exponential terms in spectrum decomposition is fixed at $M=9$.
The imaginary part of $G^{r}_{\uparrow \uparrow}(t>0)$ has the same convergence
behavior and is ignored here.

Obviously the short-time dynamics converges more easily than long-time dynamics
when the truncation level $L$ increases. This is consistent with the frequency-domain
convergence features with respect to truncation level indicated in \Fig{figSM2}.
This is because that dynamical properties at high and far-from-resonance frequency
range are dominated by the short-time evolution.

As illustrated in the inset of \Fig{figSM3}, the real time evolution has a long-time
oscillatory tail. To obtain quantitative accurate spectral function at high and
off-resonant frequencies, one only need evolve relative short time $t$. In contrast,
the accurate Kondo peaks can be achieved only when long enough oscillatory evolution
is included. Therefore the long-time oscillatory tail is crucial for the Kondo
resonance peaks. It can be viewed as another evidence of the existence of Kondo
correlation.

Therefore for strongly correlated quantum impurity systems at low
temperatures, where Kondo signatures appear at the position of chemical
potentials $\mu_{\alpha}$, one must propagate $G^{r}_{\uparrow
\uparrow}(t>0)$ to a long enough time to include the significant
long-time memory of reservoir. Consequently, it is very time-consuming
to obtain quantitative accuracy at the frequencies of Kondo peaks. For
the present parameters in the main text, when $V_{0}=0.7\,$meV, $L=4$
and $M=9$, the time evolution needs to be extended over $190\,$ps
(costing $10000$ minutes of CPU time) to obtain accurate Kondo peaks.

\begin{figure}
\includegraphics[width=0.82\columnwidth]{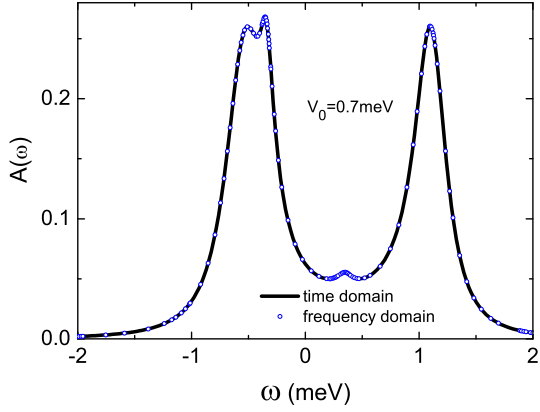}
\caption{
HEOM calculated spectral function $A(\omega)$ of SIAM obtained in both time domain
with $t=190\,$ps and frequency domain, in unit of $(\pi\Gamma)^{-1}$. The parameters
are the same as those specified in \Fig{figSM3}. The truncation level is fixed at
$L=4$ and the number of exponential terms in spectrum decomposition is fixed at $M=9$.
}
\label{figSM4}
\end{figure}

\begin{figure}
\includegraphics[width=0.8\columnwidth]{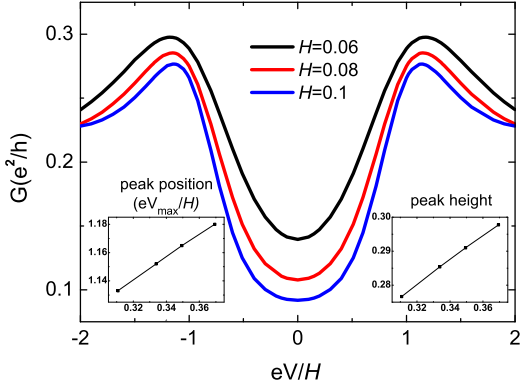}
\caption{
Calculated differential conductance $G=dI/dV$ of a symmetrical SIAM
under magnetic fields of different magnitudes ($H$). The other parameters
adopted are (in meV): $\Gamma=\Gamma_{\rm L}=\Gamma_{\rm R}=0.1$,
$\epsilon=\epsilon_{\uparrow}=\epsilon_{\downarrow}=-0.6$, $U=1.2$,
$W_{\rm L}=W_{\rm R}=2$, and $T = 0.01$. The insets display the variations
of peak position $eV_{\rm max}/H$ and peak height with respect to
$[\ln(H/(2T_{K}) \sqrt{\pi e/2})]^{-1}$, respectively. Here, the Kondo
temperature is evaluated as $T_K = \sqrt{\Gamma U/2}
\,e^{\,\pi[\epsilon (\epsilon+U) + \Gamma^2]/(2\Gamma U)} \approx 0.002508\,$meV.
}
\label{figSM5}
\end{figure}

An alternative numerical scheme is carried out in frequency domain.
In frequency-domain scheme one solves the following sparse linear
algebra problem at a fixed frequency $\omega$,
\begin{align}
&-i\omega\tilde{\rho}^{(n)}_{j_1\cdots j_n}(\omega)-\tilde{\rho}^{(n)}_{j_1\cdots j_n}(0)=
-\big[ i\mathcal{L}_{s}+\sum_{r=1}^{n}\gamma_{j_r}\big]\tilde{\rho}^{(n)}_{j_1\cdots j_n}(\omega)
\nl &
-i \sideset{}{'}\sum_{j}{\cal A}_{\bar j}\, \tilde{\rho}^{(n+1)}_{j_{1}\cdots j_{n} j}(\omega)
-i \sum_{r=1}^{n}(-)^{n-r}\, {\cal C}_{j_r}\,
\tilde{\rho}^{(n-1)}_{j_{1}\cdots j_{r-1}j_{r+1}\cdots j_{n}}(\omega)\, .
\end{align}
where $\tilde{\rho}^{(n)}_{j_1\cdots j_n}(\omega)$ is the half-Fourier
transformation of $\tilde{\rho}^{(n)}_{j_1\cdots j_n}(t)$. And
$\tilde{\rho}^{(n)}_{j_1\cdots j_n}(0)$ is the initial condition of the
corresponding correlation/response function. To obtain the whole spectrum,
a full scan for the whole frequency domain is needed.

The equivalence between the time- and frequency-domain schemes is exemplified
with \Fig{figSM4}, where the spectral function $A(\omega)$ of a SIAM is calculated
with both time- and frequency-domain schemes. Apparently, the results of both
schemes agree perfectly with each other.

In practice, one can employ a hybrid time- and frequency-domain scheme.
The overall profile can be plotted with the time-domain scheme, while the
detailed structures of the resonance and Kondo peaks can be exploited with the
frequency-domain scheme. This hybrid scheme provides an accurate and efficient
method to calculate the full spectrum of a strongly correlated quantum impurity
system in the Kondo regime.

We then investigate the effects of external magnetic field on the
static differential conductance of an SIAM system in the Kondo regime.
The magnetic field causes the Zeeman splitting for the impurity level
through an additional interaction Hamiltonian of $\hat{H}_{\rm M} = - H
\hat{S}_z$.
The effects of the magnetic field on the nonequilibrium properties of
the SIAM, such as the steady-state differential conductance ($G \equiv
dI/dV$), have been explored in the literature.\cite{Mei932601,Kon01236801,
Kon02125304} Konik, Saleur, and Ludwig\cite{Kon02125304} have analyzed
the asymptotic behavior of the peak position ($eV_{\rm max}/H$) and peak
height with increasing $H$ --they vary logarithmically in $H$. Taking
the peak position as an example, it has been found that as $H/T_{K}
\rightarrow \infty$,\cite{Kon02125304}
\be \label{evmax}
  \frac{eV_{\rm max}}{H} \propto \left[\ln\left( \frac{H}{2T_{K}}
  \sqrt{\frac{\pi e}{2}} \right)\right]^{-1}.
\ee

Figure~\ref{figSM5} depicts the calculated $G$ versus $eV/H$ for
various values of $H (\gg T_{K})$. Here, $\mu_L$ is fixed at its
equilibrium value, while $\mu_R$ is shifted by the applied bias
voltage. In contrast to Ref.\,\onlinecite{Kon02125304} where zero
temperature is considered, the HEOM calculations are done at a low but
finite temperature, $T = 0.01\,$meV, because of our limited
computational resources. The insets of \Fig{figSM5} show clearly that
both $eV_{\rm max}/H$ and peak height indeed vary linearly with respect
to the quantity at the right-hand side of \Eq{evmax}. Moreover, by
linear extrapolation to $H \rightarrow \infty$, the ratio $eV_{\rm
max}/H$ approaches a value of around $0.88$, smaller than $1$.
Therefore, different from the $T = 0$ case where $eV_{\rm max}/H$ is
always smaller than $1$, our calculations indicate that, at a finite
temperature such a ratio may decrease from above $1$ to a value smaller
than $1$.
Such a trend has been observed experimentally; see for instance,
Fig.\,4(b) of Ref.\,\onlinecite{Cro98540}.


In summary, the truncation level and the number of exponential terms in spectrum
decomposition are continuously increased until convergent numerical results are
obtained. Therefore the proposed HEOM nonequilibrium linear response theory provides
an quantitatively accurate numerical tool for arbitrary response and correlation
functions of local impurity systems, and transport related response properties.
In the frame of our HEOM formalism, one can extract dynamical properties at high
and off-resonant frequency range at relative lower truncation level and relative
shorter evolution time than those at resonance frequencies, maintaining quantitative
accuracy. The hybrid time- and frequency-domain scheme provides an accurate and
efficient method to obtain the full spectrum of a quantum impurity system in the
Kondo regime.


\begin{thebibliography}{10}

\bibitem{Elz04431}
J.~M. Elzerman, R.~Hanson, L.~H. {Willems van Beveren}, B.~Witkamp, L.~M.~K. Vandersypen, and L.~P. Kouwenhoven,
\newblock Nature {\bf 430}, 431 (2004).

\bibitem{Kop06766}
F.~H.~L. Koppens, C. Buizert, K. J. Tielrooij, I. T. Vink, K. C. Nowack, T. Meunier, L. P. Kouwenhoven, and L. M. K. Vandersypen,
\newblock Nature {\bf 442}, 766 (2006).

\bibitem{Han081043}
R.~Hanson and D.~D. Awschalom,
\newblock Nature {\bf 453}, 1043 (2008).

\bibitem{Gab06499}
J.~Gabelli, G.~F\`{e}ve, J.~-M. Berroir, B.~Pla\c{c}ais, A.~Cavanna, B.~Etienne, Y.~Jin, and D.~C. Glattli,
\newblock Science {\bf 313}, 499 (2006).

\bibitem{Fev071169}
G.~F\`{e}ve, A.~Mah\'{e}, J.~-M. Berroir, T.~Kontos, B.~Pla\c{c}ais, D.~C. Glattli, A.~Cavanna, B.~Etienne, and Y.~Jin,
\newblock Science {\bf 316}, 1169 (2007).

\bibitem{Met89324}
W.~Metzner and D.~Vollhardt,
\newblock Phys. Rev. Lett. {\bf 62}, 324 (1989).

\bibitem{Geo926479}
A.~Georges and G.~Kotliar,
\newblock Phys. Rev. B {\bf 45}, 6479 (1992).

\bibitem{Geo9613}
A.~Georges, G.~Kotliar, W.~Krauth, and M.~J. Rozenberg,
\newblock Rev. Mod. Phys. {\bf 68}, 13 (1996).

\bibitem{Cro98540}
S.~M. Cronenwett, T.~H. Oosterkamp, and L.~P. Kouwenhoven,
\newblock Science {\bf 281}, 540 (1998).

\bibitem{Gol98156}
D.~Goldhaber-Gordon, H.~Shtrikman, D.~Mahalu, D.~Abusch-Magder, U.~Meirav, and M.~A. Kastner,
\newblock Nature {\bf 391}, 156 (1998).

\bibitem{Bul08395}
R.~Bulla, T.~A. Costi, and T.~Pruschke,
\newblock Rev. Mod. Phys. {\bf 80}, 395 (2008).

\bibitem{Geo921240}
A.~Georges and W.~Krauth,
\newblock Phys. Rev. Lett. {\bf 69}, 1240 (1992).

\bibitem{Ima981039}
M.~Imada, A.~Fujimori, and Y.~Tokura,
\newblock Rev. Mod. Phys. {\bf 70}, 1039 (1998).

\bibitem{Bul01045103}
R.~Bulla, T.~A. Costi, and D.~Vollhardt,
\newblock Phys. Rev. B {\bf 64}, 045103 (2001).

\bibitem{Eme872794}
V.~J. Emery,
\newblock Phys. Rev. Lett. {\bf 58}, 2794 (1987).

\bibitem{Yan031}
Y.~Yanase, T.~Jujo, T.~Nomura, H.~Ikeda, T.~Hotta, and K.~Yamada,
\newblock Phys. Rep. {\bf 387}, 1 (2003).

\bibitem{Mai05237001}
T.~A. Maier, M.~Jarrell, T.~C. Schulthess, P.~R.~C. Kent, and J.~B. White,
\newblock Phys. Rev. Lett. {\bf 95}, 237001 (2005).

\bibitem{Wil75773}
K.~G. Wilson,
\newblock Rev. Mod. Phys. {\bf 47}, 773 (1975).

\bibitem{Hof001508}
W.~Hofstetter,
\newblock Phys. Rev. Lett. {\bf 85}, 1508 (2000).

\bibitem{Whi922863}
S.~R. White,
\newblock Phys. Rev. Lett. {\bf 69}, 2863 (1992).

\bibitem{Jec02045114}
E.~Jeckelmann,
\newblock Phys. Rev. B {\bf 66}, 045114 (2002).

\bibitem{Nis04613}
S.~Nishimoto and E.~Jeckelmann,
\newblock J. Phys.: Condens. Matter {\bf 16}, 613 (2004).

\bibitem{Hir862521}
J.~E. Hirsch and R.~M. Fye,
\newblock Phys. Rev. Lett. {\bf 56}, 2521 (1986).

\bibitem{Sil902380}
R.~N. Silver, D.~S. Sivia, and J.~E. Gubernatis,
\newblock Phys. Rev. B {\bf 41}, 2380 (1990).

\bibitem{Gub916011}
J.~E. Gubernatis, M.~Jarrell, R.~N. Silver, and D.~S. Sivia,
\newblock Phys. Rev. B {\bf 44}, 6011 (1991).

\bibitem{Gul11349}
E.~Gull, A.~J. Millis, A.~I. Lichtenstein, A.~N. Rubtsov, M.~Troyer, and P.~Werner,
\newblock Rev. Mod. Phys. {\bf 83}, 349 (2011).

\bibitem{Doy06245326}
B.~Doyon and N.~Andrei,
\newblock Phys. Rev. B {\bf 73}, 245326 (2006).

\bibitem{Meh08086804}
P.~Mehta and N.~Andrei,
\newblock Phys. Rev. Lett. {\bf 100}, 086804 (2008).

\bibitem{Bou08140601}
E.~Boulat, H.~Saleur, and P.~Schmitteckert,
\newblock Phys. Rev. Lett. {\bf 101}, 140601 (2008).

\bibitem{Chu09216803}
C.-H. Chung, K.~{Le~Hur}, M.~Vojta, and P.~{W\"{o}lfle},
\newblock Phys. Rev. Lett. {\bf 102}, 216803 (2009).

\bibitem{Cos973003}
T.~A. Costi,
\newblock Phys. Rev. B {\bf 55}, 3003 (1997).

\bibitem{And05196801}
F.~B. Anders and A.~Schiller,
\newblock Phys. Rev. Lett. {\bf 95}, 196801 (2005).

\bibitem{And08066804}
F.~B. Anders,
\newblock Phys. Rev. Lett. {\bf 101}, 066804 (2008).

\bibitem{Whi04076401}
S.~R. White and A.~E. Feiguin,
\newblock Phys. Rev. Lett. {\bf 93}, 076401 (2004).

\bibitem{Jak07150603}
S.~G. Jakobs, V.~Meden, and H.~Schoeller,
\newblock Phys. Rev. Lett. {\bf 99}, 150603 (2007).

\bibitem{Gez07045324}
R.~Gezzi, T.~Pruschke, and V.~Meden,
\newblock Phys. Rev. B {\bf 75}, 045324 (2007).

\bibitem{Han99236808}
J.~E. Han and R.~J. Heary,
\newblock Phys. Rev. Lett. {\bf 99}, 236808 (2007).

\bibitem{Wer09035320}
P.~Werner, T.~Oka, and A.~J. Millis,
\newblock Phys. Rev. B {\bf 79}, 035320 (2009).

\bibitem{Sch09153302}
M.~Schir\'{o} and M.~Fabrizio,
\newblock Phys. Rev. B {\bf 79}, 153302 (2009).

\bibitem{Wei08195316}
S.~Weiss, J.~Eckel, M.~Thorwart, and R.~Egger,
\newblock Phys. Rev. B {\bf 77}, 195316 (2008).

\bibitem{Seg10205323}
D.~Segal, A.~J. Millis, and D.~R. Reichman,
\newblock Phys. Rev. B {\bf 82}, 205323 (2010).

\bibitem{Kon01236801}
R.~M. Konik, H.~Saleur, and A.~W.~W. Ludwig,
\newblock Phys. Rev. Lett. {\bf 87}, 236801 (2001).

\bibitem{Meh06216802}
P.~Mehta and N.~Andrei,
\newblock Phys. Rev. Lett. {\bf 96}, 216802 (2006).

\bibitem{Cha11195314}
S.-P. Chao and G.~Palacios,
\newblock Phys. Rev. B {\bf 83}, 195314 (2011).

\bibitem{Jin08234703}
J.~S. Jin, X.~Zheng, and Y.~J. Yan,
\newblock J. Chem. Phys. {\bf 128}, 234703 (2008).

\bibitem{Zhe09164708}
X.~Zheng, J.~S. Jin, S.~Welack, M.~Luo, and Y.~J. Yan,
\newblock J. Chem. Phys. {\bf 130}, 164708 (2009).

\bibitem{Zhe121129}
Zheng Xiao, Xu Ruixue, Xu Jian, Jin Jinshuang, Hu Jie, and Yan Yijing,
\newblock Prog. Chem. {\bf 24}, 1129 (2012).

\bibitem{Li12266403}
Z.~H. Li, N.~H. Tong, X.~Zheng, D.~Hou, J.~H. Wei, J.~Hu, and Y.~J. Yan,
\newblock Phys. Rev. Lett. {\bf 109}, 266403 (2012).

\bibitem{Ste04195318}
G.~Stefanucci and C.-O. Almbladh,
\newblock Phys. Rev. B {\bf 69}, 195318 (2004).

\bibitem{Mac06085324}
J.~Maciejko, J.~Wang, and H.~Guo,
\newblock Phys. Rev. B {\bf 74}, 085324 (2006).

\bibitem{Zhe08184112}
X.~Zheng, J.~S. Jin, and Y.~J. Yan,
\newblock J. Chem. Phys. {\bf 129}, 184112 (2008).

\bibitem{Zhe08093016}
X.~Zheng, J.~S. Jin, and Y.~J. Yan,
\newblock New J. Phys. {\bf 10}, 093016 (2008).

\bibitem{Jia12245427}
F.~Jiang, J.~S. Jin, S.~K. Wang, and Y.~J. Yan,
\newblock Phys. Rev. B {\bf 85}, 245427 (2012).

\bibitem{Cro09073102}
A.~Croy and U.~Saalmann,
\newblock Phys. Rev. B {\bf 80}, 073102 (2009).

\bibitem{Hu10101106}
J.~Hu, R.~X. Xu, and Y.~J. Yan,
\newblock J. Chem. Phys. {\bf 133}, 101106 (2010).

\bibitem{Hu11244106}
J.~Hu, M.~Luo, F.~Jiang, R.~X. Xu, and Y.~J. Yan,
\newblock J. Chem. Phys. {\bf 134}, 244106 (2011).

\bibitem{Zhe09124508}
X.~Zheng, J.~Y. Luo, J.~S. Jin, and Y.~J. Yan,
\newblock J. Chem. Phys. {\bf 130}, 124508 (2009).

\bibitem{Mo05084115}
Y.~Mo, R.~X. Xu, P.~Cui, and Y.~J. Yan,
\newblock J. Chem. Phys. {\bf 122}, 084115 (2005).

\bibitem{Zhu115678}
K.~B. Zhu, R.~X. Xu, H.~Y. Zhang, J.~Hu, and Y.~J. Yan,
\newblock J. Phys. Chem. B {\bf 115}, 5678 (2011).

\bibitem{Xu11497}
J.~Xu, R.~X. Xu, D.~Abramavicius, H.~D. Zhang, and Y.~J. Yan,
\newblock Chin. J. Chem. Phys. {\bf 24}, 497 (2011).

\bibitem{Xu13024106}
J.~Xu, H.~D. Zhang, R.~X. Xu, and Y.~J. Yan,
\newblock J. Chem. Phys. {\bf 138}, 024106 (2013).

\bibitem{Xu05041103}
R.~X. Xu, P.~Cui, X.~Q. Li, Y.~Mo, and Y.~J. Yan,
\newblock J. Chem. Phys. {\bf 122}, 041103 (2005).

\bibitem{Shi09084105}
Q.~Shi, L.~P. Chen, G.~J. Nan, R.~X. Xu, and Y.~J. Yan,
\newblock J. Chem. Phys. {\bf 130}, 084105 (2009).

\bibitem{Bak96}
G.~A. {Baker Jr.} and P.~Graves-Morris,
\newblock {\em Pad\'{e} Approximants},
\newblock Cambridge University Press, New York, 1996,
\newblock 2nd ed.

\bibitem{SM_skw}
See the Supplemental Material for the numerical details on the HEOM evaluation
  of nonequilibrium dynamical properties of quantum impurity systems.

\bibitem{Hew93}
A.~C. Hewson,
\newblock {\em The Kondo Problem to Heavy Fermions},
\newblock Cambridge University Press, Cambridge, 1993.

\bibitem{Agu04206601}
R.~Aguado and T.~Brandes,
\newblock Phys. Rev. Lett. {\bf 92}, 206601 (2004).

\bibitem{Fli05411}
C.~Flindt, T.~{Novotn\'{y}}, and A.-P. Jauho,
\newblock Physica E {\bf 29}, 411 (2005).

\bibitem{Luo07085325}
J.~Y. Luo, X.~Q. Li, and Y.~J. Yan,
\newblock Phys. Rev. B {\bf 76}, 085325 (2007).

\bibitem{Mer12075153}
L.~Merker, A.~Weichselbaum, and T.~A. Costi,
\newblock Phys. Rev. B {\bf 86}, 075153 (2012).

\bibitem{But93364}
M.~{B\"{u}ttiker}, H.~Thomas, and A.~{Pr\^{e}tre},
\newblock Phys. Lett. A {\bf 180}, 364 (1993).

\bibitem{But934114}
M.~{B\"{u}ttiker}, A.~{Pr\^{e}tre}, and H.~Thomas,
\newblock Phys. Rev. Lett. {\bf 70}, 4114 (1993).

\bibitem{Pre968130}
A.~{Pr\^{e}tre}, H.~Thomas, and M.~{B\"{u}ttiker},
\newblock Phys. Rev. B {\bf 54}, 8130 (1996).

\bibitem{But964793}
M.~{B\"uttiker},
\newblock J. Math. Phys. {\bf 37}, 4793 (1996).

\bibitem{Win938487}
N.~S. Wingreen, A.-P. Jauho, and Y.~Meir,
\newblock Phys. Rev. B {\bf 48}, 8487 (1993).

\bibitem{Jau945528}
A.-P. Jauho, N.~S. Wingreen, and Y.~Meir,
\newblock Phys. Rev. B {\bf 50}, 5528 (1994).

\bibitem{Ana957632}
M.~P. Anantram and S.~Datta,
\newblock Phys. Rev. B {\bf 51}, 7632 (1995).

\bibitem{Nig06206804}
S.~E. Nigg, R.~{L\'{o}pez}, and M.~{B\"{u}ttiker},
\newblock Phys. Rev. Lett. {\bf 97}, 206804 (2006).

\bibitem{Wan07155336}
J.~Wang, B.~Wang, and H.~Guo,
\newblock Phys. Rev. B {\bf 75}, 155336 (2007).

\bibitem{Kam008154}
A.~Kaminski, Y.~V. Nazarov, and L.~I. Glazman,
\newblock Phys. Rev. B {\bf 62}, 8154 (2000).

\bibitem{Ros01156802}
A.~Rosch, J.~Kroha, and P.~W\"olfle,
\newblock Phys. Rev. Lett. {\bf 87}, 156802 (2001).

\bibitem{Pus04R513}
M.~Pustilnik and L.~Glazman,
\newblock J. Phys.: Condens. Matter {\bf 16}, R153 (2004).


\bibitem{Mei932601} Y.~Meir, N.~S. Wingreen, and P.~A. Lee,
\newblock Phys. Rev. Lett. {\bf 70}, 2601 (1993).

\bibitem{Kon02125304} R.~M. Konik, H.~Saleur, and A.~Ludwig,
\newblock Phys. Rev. B {\bf 66}, 125304 (2002).


\end{thebibliography}

\end{document}